\pdfoutput=1
\documentclass[12pt,a4paper]{article}

\usepackage{subcaption}
\usepackage{booktabs}

\usepackage{ifthen} 
\newboolean{pdflatex}
\setboolean{pdflatex}{true} 

\newboolean{articletitles}
\setboolean{articletitles}{true} 

\newboolean{uprightparticles}
\setboolean{uprightparticles}{false} 

\def\paperauthors{LHCb collaboration} 
\def\paperasciititle{Search for long-lived particles decaying to $e\mu\nu$} 
\def\papertitle{Search for long-lived particles decaying to $\epm \mump \neu$} 
\def\paperkeywords{{High Energy Physics}, {LHCb}} 
\def\papercopyright{\the\year\ CERN for the benefit of the LHCb collaboration} 
\def\paperlicence{CC BY 4.0 licence}
\def\paperlicenceurl{https://creativecommons.org/licenses/by/4.0/}

\usepackage[top=1in, bottom=1.25in, left=1in, right=1in]{geometry}

\usepackage{microtype}
\usepackage{lineno}  
\usepackage{xspace} 
\usepackage{caption} 

\usepackage{graphicx}  
\usepackage{color}
\usepackage{colortbl}
\graphicspath{{./figs/}} 

\usepackage{amsmath} 
\usepackage{amssymb}
\usepackage{amsfonts}
\usepackage{upgreek} 

\newcommand*\patchAmsMathEnvironmentForLineno[1]{%
\expandafter\let\csname old#1\expandafter\endcsname\csname #1\endcsname
\expandafter\let\csname oldend#1\expandafter\endcsname\csname
end#1\endcsname
 \renewenvironment{#1}%
   {\linenomath\csname old#1\endcsname}%
   {\csname oldend#1\endcsname\endlinenomath}%
}
\newcommand*\patchBothAmsMathEnvironmentsForLineno[1]{%
  \patchAmsMathEnvironmentForLineno{#1}%
  \patchAmsMathEnvironmentForLineno{#1*}%
}
\AtBeginDocument{%
\patchBothAmsMathEnvironmentsForLineno{equation}%
\patchBothAmsMathEnvironmentsForLineno{align}%
\patchBothAmsMathEnvironmentsForLineno{flalign}%
\patchBothAmsMathEnvironmentsForLineno{alignat}%
\patchBothAmsMathEnvironmentsForLineno{gather}%
\patchBothAmsMathEnvironmentsForLineno{multline}%
\patchBothAmsMathEnvironmentsForLineno{eqnarray}%
}

\usepackage{hyperxmp}

\usepackage[pdftex,
            pdfauthor={\paperauthors},
            pdftitle={\paperasciititle},
            pdfkeywords={\paperkeywords},
            pdfcopyright={Copyright (C) \papercopyright},
            pdflicenseurl={\paperlicenceurl}]{hyperref}

\usepackage[colorinlistoftodos,textsize=scriptsize]{todonotes}

\usepackage[bottom,flushmargin,hang,multiple]{footmisc}

\usepackage[all]{hypcap} 

\usepackage{xspace} 
\usepackage{upgreek}

\def\lhcb   {\mbox{LHCb}\xspace}
\def\atlas  {\mbox{ATLAS}\xspace}
\def\cms    {\mbox{CMS}\xspace}

\def\lhc    {\mbox{LHC}\xspace}

\def\velo   {VELO\xspace}

\def\ecal   {ECAL\xspace}
\def\hcal   {HCAL\xspace}
\def\MagUp {\mbox{\em Mag\kern -0.05em Up}\xspace}

\ifthenelse{\boolean{uprightparticles}}%
{

 \def\Pmu         {\ensuremath{\upmu}\xspace}                 
 \def\Pnu         {\ensuremath{\upnu}\xspace}

 \def\Pchi        {\ensuremath{\upchi}\xspace}                 
 \def\Ppsi        {\ensuremath{\uppsi}\xspace}

 \def\PDelta      {\ensuremath{\Delta}\xspace}                 
 \def\PXi         {\ensuremath{\Xi}\xspace}                 
 \def\PLambda     {\ensuremath{\Lambda}\xspace}                 
 \def\PSigma      {\ensuremath{\Sigma}\xspace}                 
 \def\POmega      {\ensuremath{\Omega}\xspace}                 
 \def\PUpsilon    {\ensuremath{\Upsilon}\xspace}

 \def\PB      {\ensuremath{\mathrm{B}}\xspace}                 
                  
 \def\PD      {\ensuremath{\mathrm{D}}\xspace}

 \def\PH      {\ensuremath{\mathrm{H}}\xspace}                 
                  
 \def\PJ      {\ensuremath{\mathrm{J}}\xspace}                 
 \def\PK      {\ensuremath{\mathrm{K}}\xspace}

 \def\PW      {\ensuremath{\mathrm{W}}\xspace}

 \def\PZ      {\ensuremath{\mathrm{Z}}\xspace}                 
                  
 \def\Pb      {\ensuremath{\mathrm{b}}\xspace}                 
 \def\Pc      {\ensuremath{\mathrm{c}}\xspace}                 
                  
 \def\Pe      {\ensuremath{\mathrm{e}}\xspace}

 \def\Pi      {\ensuremath{\mathrm{i}}\xspace}

 \def\Pp      {\ensuremath{\mathrm{p}}\xspace}                 
 \def\Pq      {\ensuremath{\mathrm{q}}\xspace}                 
                  
 \def\Ps      {\ensuremath{\mathrm{s}}\xspace}

 \def\thebaroffset{0.0em}
}
{

 \def\Pmu         {\ensuremath{\mu}\xspace}                 
 \def\Pnu         {\ensuremath{\nu}\xspace}

 \def\Pchi        {\ensuremath{\chi}\xspace}                 
 \def\Ppsi        {\ensuremath{\psi}\xspace}                 
                  
 \mathchardef\PDelta="7101
 \mathchardef\PXi="7104
 \mathchardef\PLambda="7103
 \mathchardef\PSigma="7106
 \mathchardef\POmega="710A
 \mathchardef\PUpsilon="7107
                  
 \def\PB      {\ensuremath{B}\xspace}                 
                  
 \def\PD      {\ensuremath{D}\xspace}

 \def\PH      {\ensuremath{H}\xspace}                 
                  
 \def\PJ      {\ensuremath{J}\xspace}                 
 \def\PK      {\ensuremath{K}\xspace}

 \def\PW      {\ensuremath{W}\xspace}

 \def\PZ      {\ensuremath{Z}\xspace}                 
                  
 \def\Pb      {\ensuremath{b}\xspace}                 
 \def\Pc      {\ensuremath{c}\xspace}                 
                  
 \def\Pe      {\ensuremath{e}\xspace}

 \def\Pi      {\ensuremath{i}\xspace}

 \def\Pp      {\ensuremath{p}\xspace}                 
 \def\Pq      {\ensuremath{q}\xspace}                 
                  
 \def\Ps      {\ensuremath{s}\xspace}

 \def\thebaroffset{0.18em}
}
\newcommand{\offsetoverline}[2][\thebaroffset]{\kern #1\overline{\kern -#1 #2}}%

\makeatletter
\ifcase \@ptsize \relax
  \newcommand{\miniscule}{\@setfontsize\miniscule{4}{5}}
\or
  \newcommand{\miniscule}{\@setfontsize\miniscule{5}{6}}
\or
  \newcommand{\miniscule}{\@setfontsize\miniscule{5}{6}}
\fi
\makeatother

\DeclareRobustCommand{\optbar}[1]{\shortstack{{\miniscule (\rule[.5ex]{1.25em}{.18mm})}
  \\ [-.7ex] $#1$}}

\def\electron   {{\ensuremath{\Pe}}\xspace}

\def\epm        {{\ensuremath{\Pe^\pm}}\xspace} 
 
\def\epem       {{\ensuremath{\Pe^+\Pe^-}}\xspace}

\def\muon       {{\ensuremath{\Pmu}}\xspace}

\def\mupm       {{\ensuremath{\Pmu^\pm}}\xspace} 
\def\mump       {{\ensuremath{\Pmu^\mp}}\xspace} 
\def\mumu       {{\ensuremath{\Pmu^+\Pmu^-}}\xspace}

\def\neu        {{\ensuremath{\Pnu}}\xspace}

\def\H      {{\ensuremath{\PH^0}}\xspace}

\def\W      {{\ensuremath{\PW}}\xspace}

\def\Z      {{\ensuremath{\PZ}}\xspace}

\def\quark     {{\ensuremath{\Pq}}\xspace}
\def\quarkbar  {{\ensuremath{\overline \quark}}\xspace}

\def\squark    {{\ensuremath{\Ps}}\xspace}

\def\cquark    {{\ensuremath{\Pc}}\xspace}

\def\bquark    {{\ensuremath{\Pb}}\xspace}
\def\bquarkbar {{\ensuremath{\overline \bquark}}\xspace}
\def\bbbar     {{\ensuremath{\bquark\bquarkbar}}\xspace}

\def\KorKbar {\kern \thebaroffset\optbar{\kern -\thebaroffset \PK}{}\xspace}

\def\D       {{\ensuremath{\PD}}\xspace}

\def\DorDbar {\kern \thebaroffset\optbar{\kern -\thebaroffset \PD}\xspace}

\def\Dp      {{\ensuremath{\D^+}}\xspace}
\def\Dm      {{\ensuremath{\D^-}}\xspace}

\def\DpDm    {\ensuremath{\Dp {\kern -0.16em \Dm}}\xspace}

\def\B       {{\ensuremath{\PB}}\xspace}

\def\BorBbar {\kern \thebaroffset\optbar{\kern -\thebaroffset \PB}\xspace}

\def\Bd      {{\ensuremath{\B^0}}\xspace}

\def\BdorBdbar {\kern \thebaroffset\optbar{\kern -\thebaroffset \Bd}\xspace}

\def\Bs      {{\ensuremath{\B^0_\squark}}\xspace}

\def\BsorBsbar {\kern \thebaroffset\optbar{\kern -\thebaroffset \Bs}\xspace}

\def\jpsi     {{\ensuremath{{\PJ\mskip -3mu/\mskip -2mu\Ppsi}}}\xspace}

\def\Upsilonres  {{\ensuremath{\PUpsilon}}\xspace}
\def\Y#1S{\ensuremath{\PUpsilon{(#1S)}}\xspace}

\def\proton      {{\ensuremath{\Pp}}\xspace}

\def\LorLbar     {\kern \thebaroffset\optbar{\kern -\thebaroffset \PLambda}\xspace}

\newcommand{\decay}[2]{\ensuremath{#1\!\to #2}\xspace} 

\def\to                 {\ensuremath{\rightarrow}\xspace}

\def\AT#1     {\ensuremath{A_{\mathrm{T}}^{#1}}\xspace}           

\def\C#1      {\ensuremath{\mathcal{C}_{#1}}\xspace}                       
\def\Cp#1     {\ensuremath{\mathcal{C}_{#1}^{'}}\xspace}                    
\def\Ceff#1   {\ensuremath{\mathcal{C}_{#1}^{\mathrm{(eff)}}}\xspace}        
\def\Cpeff#1  {\ensuremath{\mathcal{C}_{#1}^{'\mathrm{(eff)}}}\xspace}       
\def\Ope#1    {\ensuremath{\mathcal{O}_{#1}}\xspace}                       
\def\Opep#1   {\ensuremath{\mathcal{O}_{#1}^{'}}\xspace}                    

\newcommand{\nospaceunit}[1]{\ensuremath{\text{#1}}}       
\newcommand{\aunit}[1]{\ensuremath{\text{\,#1}}}       

\newcommand{\tev}{\aunit{Te\kern -0.1em V}\xspace}
\newcommand{\gev}{\aunit{Ge\kern -0.1em V}\xspace}
\newcommand{\mev}{\aunit{Me\kern -0.1em V}\xspace}
\newcommand{\kev}{\aunit{ke\kern -0.1em V}\xspace}
\newcommand{\ev}{\aunit{e\kern -0.1em V}\xspace}
 
\newcommand{\mevc}{\ensuremath{\aunit{Me\kern -0.1em V\!/}c}\xspace}
\newcommand{\gevc}{\ensuremath{\aunit{Ge\kern -0.1em V\!/}c}\xspace}
\newcommand{\mevcc}{\ensuremath{\aunit{Me\kern -0.1em V\!/}c^2}\xspace}
\newcommand{\gevcc}{\ensuremath{\aunit{Ge\kern -0.1em V\!/}c^2}\xspace}

\def\mm   {\aunit{mm}\xspace}

\def\mum  {\ensuremath{\,\upmu\nospaceunit{m}}\xspace}

\def\mub{\ensuremath{\,\upmu\nospaceunit{b}}\xspace}

\def\pb {\aunit{pb}\xspace}

\def\fb   {\ensuremath{\aunit{fb}}\xspace}
\def\invfb   {\ensuremath{\fb^{-1}}\xspace}

\def\ns   {\ensuremath{\aunit{ns}}\xspace}
\def\ps   {\ensuremath{\aunit{ps}}\xspace}

\newcommand{\chisq}{\ensuremath{\chi^2}\xspace}

\newcommand{\chisqip}{\ensuremath{\chi^2_{\text{IP}}}\xspace}

\def\gsim{{~\raise.15em\hbox{$>$}\kern-.85em
          \lower.35em\hbox{$\sim$}~}\xspace}
\def\lsim{{~\raise.15em\hbox{$<$}\kern-.85em
          \lower.35em\hbox{$\sim$}~}\xspace}

\def\sqs   {\ensuremath{\protect\sqrt{s}}\xspace}

\def\pt         {\ensuremath{p_{\mathrm{T}}}\xspace}

\def\ptot       {\ensuremath{p}\xspace}

\def\geant      {\mbox{\textsc{Geant4}}\xspace}

\def\pythia     {\mbox{\textsc{Pythia}}\xspace}
\def\madgraph     {\mbox{\textsc{MadGraph}}\xspace}
\def\madspin     {\mbox{\textsc{MadSpin}}\xspace}

\def\tell1  {TELL1\xspace}
\def\ukl1   {UKL1\xspace}

\def\OneChi{\ensuremath{\tilde{\Pchi}^{1}_{0}}\xspace}

\def\mLLP {\ensuremath{m_{\text{LLP}}}\xspace}
\def\tauLLP {\ensuremath{\tau_{\text{LLP}}}\xspace}
\def\LLPdec {\decay{\text{LLP}}{\epm\mump\neu}}
\def\mcorr {\ensuremath{m_{\text{corr}}}\xspace}

\def\dr{\ensuremath{\Delta R}\xspace}

\def\Zmumu{\ensuremath{\Z \to \mumu}\xspace}
\def\Zee{\ensuremath{\Z \to \epem}\xspace}
\def\Upsmumu{\ensuremath{\Upsilonres \to \mumu}\xspace}

\def\pythia8{\mbox{\textsc{Pythia8}}\xspace}

\usepackage{cite} 
\usepackage{mciteplus}

\usepackage{longtable} 
\usepackage[detect-all]{siunitx}
\begin{document}

\renewcommand{\thefootnote}{\fnsymbol{footnote}}
\setcounter{footnote}{1}


\begin{titlepage}
\pagenumbering{roman}

\vspace*{-1.5cm}
\centerline{\large EUROPEAN ORGANIZATION FOR NUCLEAR RESEARCH (CERN)}
\vspace*{1.5cm}
\noindent
\begin{tabular*}{\linewidth}{lc@{\extracolsep{\fill}}r@{\extracolsep{0pt}}}
\ifthenelse{\boolean{pdflatex}}
{\vspace*{-1.5cm}\mbox{\!\!\!\includegraphics[width=.14\textwidth]{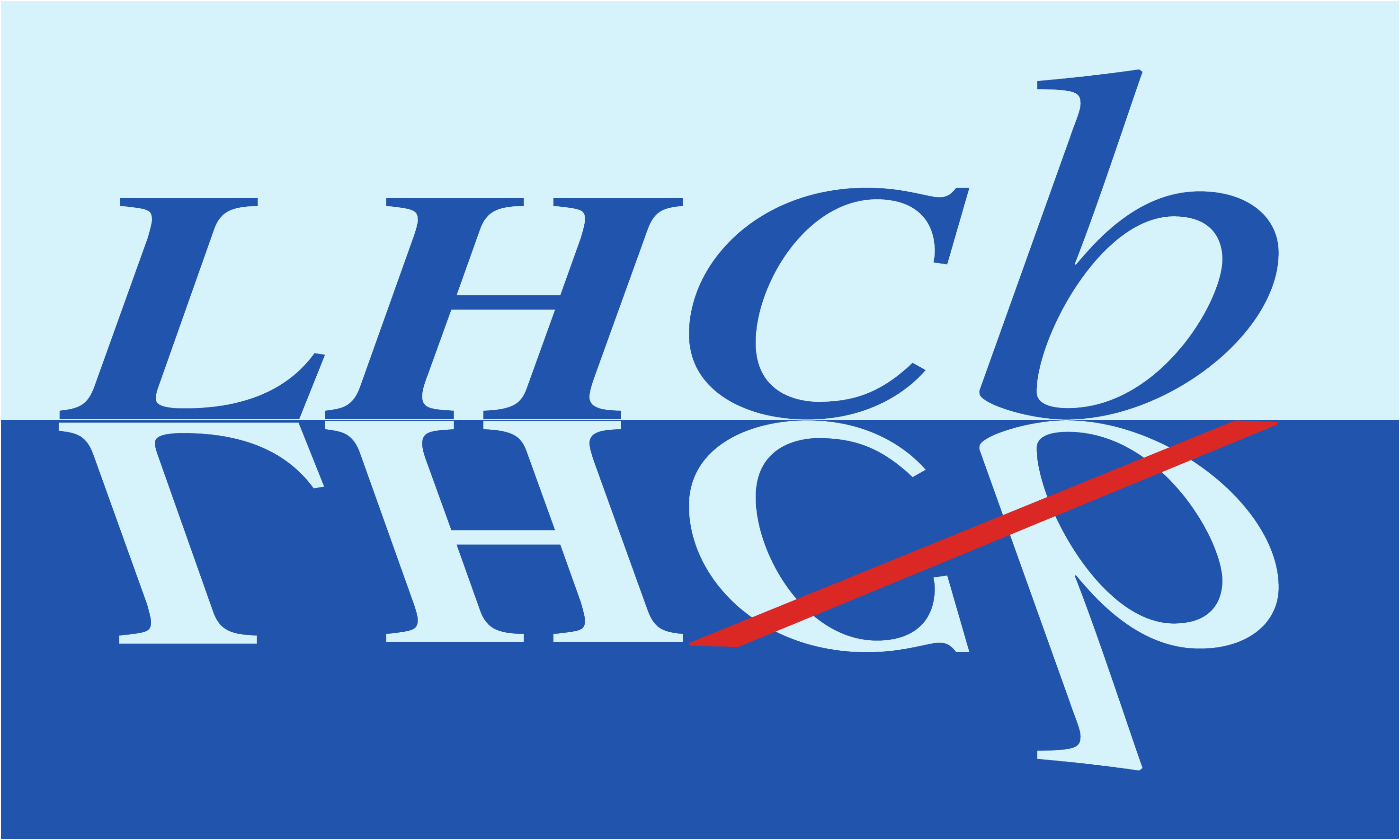}} & &}%
{\vspace*{-1.2cm}\mbox{\!\!\!\includegraphics[width=.12\textwidth]{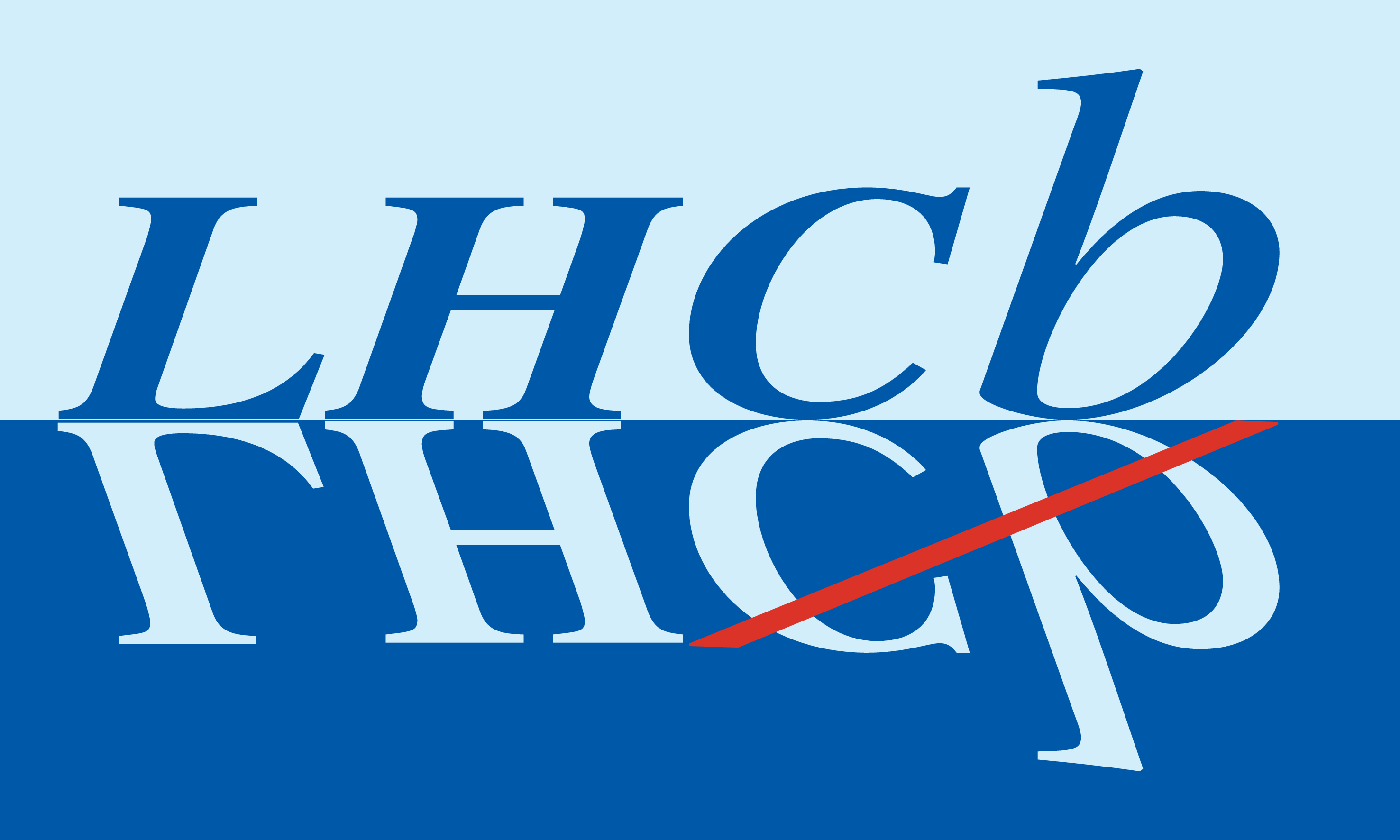}} & &}%
\\
 & & CERN-EP-2020-212 \\  
 & & LHCb-PAPER-2020-027 \\  
 & & March, 30 2021 \\ 
 & & \\
\end{tabular*}

\vspace*{4.0cm}

{\normalfont\bfseries\boldmath\huge
\begin{center}
  \papertitle 
\end{center}
}

\vspace*{2.0cm}

\begin{center}
\paperauthors\footnote{Authors are listed at the end of this paper.}
\end{center}

\vspace{\fill}

\begin{abstract}
  \noindent
Long-lived particles decaying to $\epm \mump \neu$, with masses between 7 and $50 \gevcc$ and lifetimes between 2 and $50 \ps$, are searched for by looking at displaced vertices containing electrons and muons of opposite charges. The search is performed using $5.4 \invfb$ of \proton\proton collisions collected with the \lhcb detector at a centre-of-mass energy of $\sqs = 13 \tev$.
Three mechanisms of production of long-lived particles are considered: the direct pair production from quark interactions, the pair production from the decay of a Standard-Model-like Higgs boson with a mass of $125 \gevcc$, and the charged current production from an on-shell \W boson with an additional lepton.
No evidence of these long-lived states is obtained and upper limits on the production cross-section times branching fraction are set on the different production modes.
  
\end{abstract}

\vspace*{2.0cm}

\begin{center}
  Published in Eur.~Phys.~J.~C81 (2021) 261
\end{center}

\vspace{\fill}

{\footnotesize 
\centerline{\copyright~\papercopyright. \href{\paperlicenceurl}{\paperlicence}.}}
\vspace*{2mm}

\end{titlepage}


\newpage
\setcounter{page}{2}
\mbox{~}
%
%
%
%


\renewcommand{\thefootnote}{\arabic{footnote}}
\setcounter{footnote}{0}

\cleardoublepage


\pagestyle{plain} 
\setcounter{page}{1}
\pagenumbering{arabic}


\section{Introduction}
\label{sec:Introduction}

\begin{figure}[b]
  \begin{center}
  
        \includegraphics[width=0.95\textwidth]{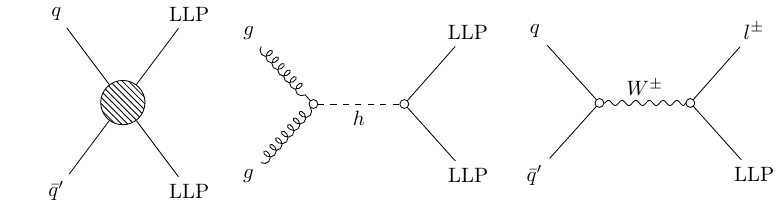}
        
  \vspace*{-0.5cm}
  \end{center}
\caption{\small Production modes of the LLP considered in this search. From left to right: direct pair production (DPP), decay of a SM-like Higgs with a mass of $125 \gevcc$ produced by gluon-gluon fusion (HIG) and production by charged current (CC).}
\label{fig:diag_prod}
\end{figure}

A variety of models beyond the Standard Model (SM) feature the existence of new massive particles with lifetimes that can be long, compared to the SM particles at the weak scale. These so-called long-lived particles (LLP) appear, for example, in Supersymmetry or extensions to the SM that predict right-handed neutrinos \cite{Alimena:2019zri}. The study presented in this paper focuses on the search for decays of neutral LLPs using three production mechanisms: direct pair production (DPP), pair production from the decay of a SM-like Higgs boson with a mass of $125 \gevcc$ (HIG), and from charged current (CC) processes. Diagrams for each production mode are shown in Fig.~\ref{fig:diag_prod}. The production of LLPs from the decay of a SM-like Higgs boson has been studied in several searches conducted by the \cms, \atlas and \lhcb experiments, using LLP decays to light-flavour jets \cite{Aad:2015asa, Aad:2015uaa, LHCb-PAPER-2016-047, LHCb-PAPER-2016-065, LHCb-PAPER-2016-014}, $b$-quark jets \cite{Aaboud:2018iil} and light leptons \cite{CMS:2014hka, Aaboud:2018jbr}.
In this study the LLP can be a neutralino \OneChi, in R-parity-violating supersymmetric models \cite{Martin:1997ns}, or a right-handed neutrino $N$ decaying to two charged leptons and a neutrino \cite{PhysRevLett.44.912, Yanagida:1980xy, GellMann:1980vs}.
Searches for \LLPdec decays have been performed by the \atlas experiment in the context of Supersymmetry \cite{Aad:2019tcc}, and also with right-handed neutrinos \cite{Aad:2019kiz}.

 The first direct \LLPdec search at the \lhcb experiment is presented in this paper. The \lhcb detector probes the forward rapidity region that is only partially covered by the other \lhc experiments, and triggers on particles with low transverse momenta, which allows the experiment to explore relatively small LLP masses.
In the present study, displaced vertices consisting of an electron and a muon of opposite charges are searched for in $\proton \proton$ collisions at a centre-of-mass energy of $\sqs = 13 \tev$, using a data sample corresponding to an integrated luminosity of $5.38 \pm 0.11 \invfb$ collected with the \lhcb detector in \numrange{2016}{2018}. The momentum of the neutrino in the final state can be partly reconstructed from the misalignment between the LLP flight direction and the momentum of the electron and muon system. The explored masses of the LLP (\mLLP) range from 7 to $50 \gevcc$ and lifetimes (\tauLLP) range from 2 to $50 \ps$.
This search enlarges the domain of searches for heavy LLPs at \lhcb, which previously probed for displaced jets \cite{LHCb-PAPER-2016-047, LHCb-PAPER-2016-065, LHCb-PAPER-2016-014} or displaced dimuons \cite{LHCb-PAPER-2019-031, LHCb-PAPER-2013-064, LHCb-PAPER-2015-036}.

\section{Detector description}

The \lhcb detector \cite{LHCb-DP-2008-001, LHCb-DP-2014-002} is a single-arm forward spectrometer covering the pseudorapidity range $2 < \eta < 5$, designed for the study of particles containing \bquark or \cquark quarks. The detector includes a high-precision tracking system consisting of a silicon-strip vertex detector surrounding the $\proton \proton$ interaction region (VELO), a large-area silicon-strip detector located upstream of a dipole magnet with a bending power of about $4 {\aunit{Tm}\xspace}$, and three stations of silicon-strip detectors and straw drift tubes, placed downstream of the magnet. The tracking system provides a measurement of momentum, \ptot, of charged particles with a relative uncertainty that varies from $0.5\%$ at low momentum to $1.0\%$ at $200\gevc$. The minimum distance of a track to a primary $\proton\proton$ collision vertex (PV), the impact parameter (IP), is measured with a resolution of $(15 + 29/\pt) \mum$, where \pt is the component of the momentum transverse to the beam axis, in \gevc. Different types of charged hadrons are distinguished using information from two ring-imaging Cherenkov detectors. Photons, electrons and hadrons are identified by a calorimeter system consisting of scintillating-pad and preshower detectors, an electromagnetic calorimeter (\ecal) and a hadronic calorimeter (\hcal). Muons are identified by a system composed of alternating layers of iron and multiwire proportional chambers. 

The online event selection is performed by a trigger, which consists of a hardware stage based on information from the calorimeter and muon systems, followed by a software stage that carries out a full event reconstruction. During data taking an alignment and calibration of the detector is performed in near real-time and used in the software trigger~\cite{LHCb-PROC-2015-011}. Events from $\proton \proton$ collisions fulfilling the muon or electron trigger are studied. At the hardware level the muon trigger requires a muon track identified by matching hits in the muon stations, for the electron trigger a cluster in the \ecal with large transverse energy deposit is required. At the software level the muon trigger selects muons with a minimum $\pt$ of $10 \gevc$, the electron trigger selects electrons with a minimum $\pt$ of $15 \gevc$.

\section{Simulation}

Simulated samples of \LLPdec events are used to design and optimise the signal selection and to estimate the detection efficiency, but also for the construction of the signal model. Parton-level events with LLPs are generated at leading order with \madgraph \cite{Alwall:2014hca} using Universal FeynRules Outputs (UFO) \cite{Degrande:2011ua} for long-lived particle searches following Ref.~\cite{Alimena:2019zri}. For the DPP and HIG mechanisms, the UFO for the minimal supersymmetric standard model with R-parity violation \cite{Martin:1997ns} is chosen, and in this framework the signal is represented by the lightest neutralino \OneChi. For the CC production the UFO of the Left-Right Symmetric model \cite{PhysRevD.10.275, Mohapatra19752558, Senjanovic19751502} is used, and here the LLP is represented by a heavy neutrino produced from an on-shell \W boson. For all three modes, the LLP is allowed to decay into an electron and a muon with opposite charges, and a neutrino. The decay of the LLP is performed through the \madspin package \cite{Artoisenet:2012st}.
The parton shower of the events is simulated with \pythia8 \cite{Sjostrand:2006za, Sjostrand:2007gs} using a specific \lhcb configuration \cite{LHCb-PROC-2010-056} and using the CTEQ6 leading-order set of parton density functions \cite{Pumplin:2002vw}. The interaction of the particles with the detector and its response are implemented using the \geant toolkit \cite{Allison:2006ve, *Agostinelli:2002hh} as described in Ref.~\cite{LHCb-PROC-2011-006}. Signal events with $\mLLP = 7, 10, 15, 20, 30, 38$ and $50 \gevcc$ and $\tauLLP = 2, 5, 10, 25$ and $50 \ps$ are generated.

Samples are also generated for background studies and cross checks, although the background estimate in this study is based on data. The most relevant background in this analysis is from \bbbar events. 
Two distinct topologies are observed with the two leptons from the same jet or from two different jets, as discussed in Section~\ref{sec:signal_yield}.
Events generated from  ${gg /\quark \quarkbar \to \bbbar}$ processes with \pythia8, with at least one muon with $\pt > 10 \gevc$ in the \lhcb acceptance are simulated and required to satisfy the muon trigger criteria.

\section{Signal selection}

The \LLPdec candidates are reconstructed from the combination of a muon and an electron candidate of opposite charges forming a good-quality vertex within the \velo detector. The following selection of the candidates is developed and optimised using the DPP samples for each pair of \mLLP and \tauLLP values. This selection is also adopted for the study of the HIG and CC processes.

The muon and electron candidates are required to have $\pt > 1.6 \gevc$ and \mbox{$\ptot > 10 \gevc$}. The measured momentum of the electron candidates is corrected for the loss of energy due to bremsstrahlung \cite{LHCB-PAPER-2013-005}.
The muon and electron need to form a good-quality vertex displaced from any PV, with a flight distance greater than 15 times its uncertainty. In addition, the lifetime of the candidate is required to be greater than $0.5 \ps$.
For the estimate of the lifetime, the Lorentz boost is calculated from the dilepton momentum, $\ptot(e\mu)$, neglecting the contribution of the neutrino. 
The mass of the candidate is obtained from the dilepton system with a correction to account for not reconstructing the neutrino. The correction is inferred from the misalignment of the dilepton reconstructed momentum and the flight direction from the PV to the decay vertex. The corrected invariant mass is computed as ${\mcorr = \sqrt{m(e\mu)^2 + \ptot(e\mu)^2 \sin^2\theta} + \ptot(e\mu) \sin\theta}$ \cite{Abe:1997sb}, where $\theta$ is the angle formed by the dilepton momentum and the LLP flight direction. Candidates with $\mcorr < 3.3 \gevcc$ are discarded.

\begin{figure}[b!]
  \begin{center}
  
        \includegraphics[width=0.95\textwidth]{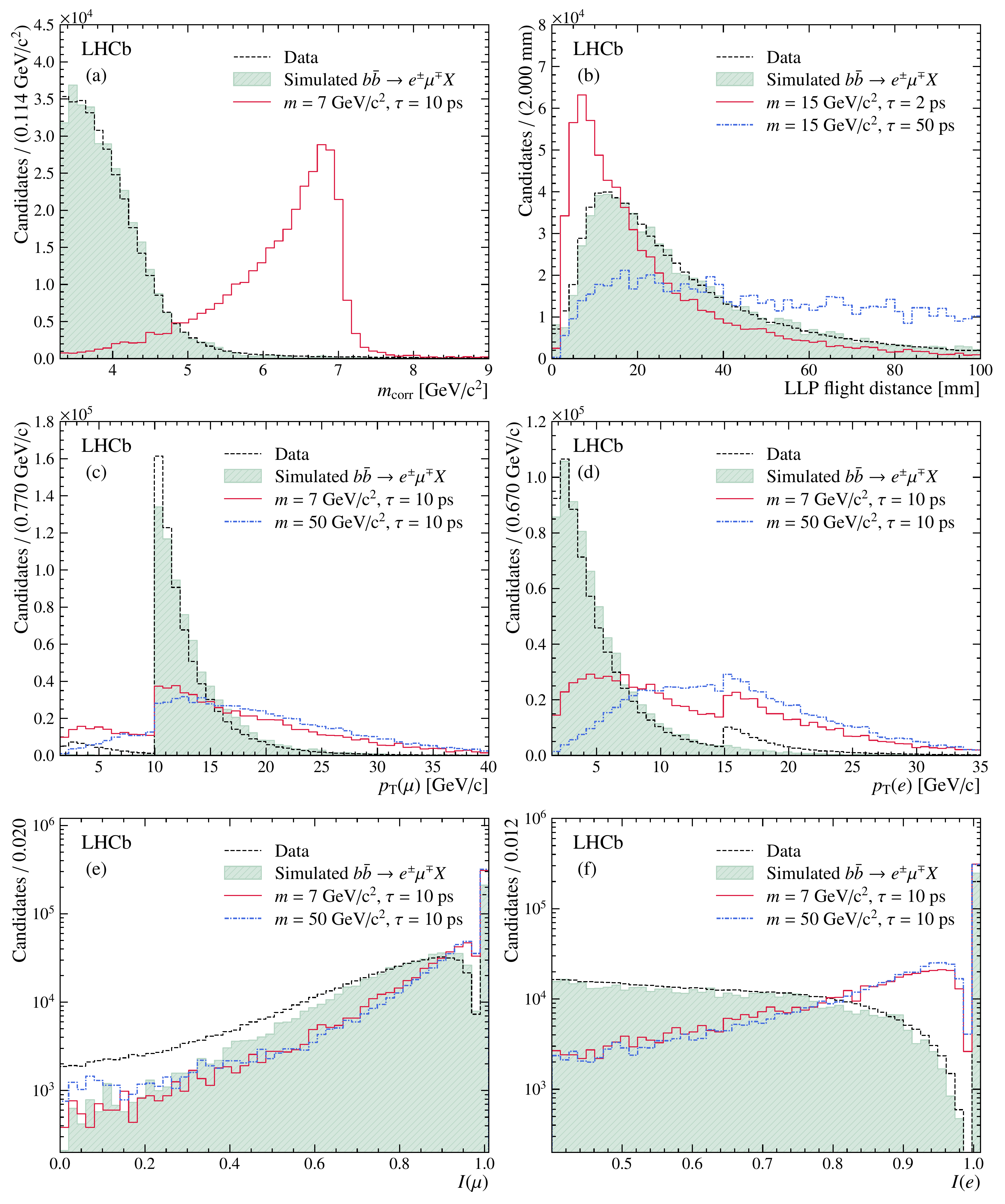}
        
  \vspace*{-0.5cm}
  \end{center}
  \caption{
    \small 
    Distributions in data (dashed black histogram) compared to simulated $\bbbar \to \epm \mump X$ (green filled histogram), showing, (a) \mcorr, (b) the LLP flight distance, (c) the transverse momentum of the muon, (d) the transverse momentum of the electron, (e) the isolation of the muon, and (f) the isolation of the electron. LLP signal distributions are also shown (coloured histograms) for different \mLLP and \tauLLP values, where the LLP is produced through the DPP mechanism. The distributions from simulation are normalised to the number of candidates in data.
    There are no simulated \bbbar candidates for $\pt(\mu) < 10 \gevcc$ due to a \pt requirement at the generation. For the same reason there is a lack of simulated \bbbar candidates for $\pt(e) > 15 \gevcc$ as candidates are required to pass the muon or electron trigger.
    }
  \label{fig:variables}
\end{figure}

To suppress the heavy-flavour background the leptons are required to be isolated from other charged particles. The isolation variable is defined as \mbox{$I = (\vec{\ptot} - \vec{\ptot}_{\text{cone}})_{\text{T}} \; / \; (\vec{\ptot} + \vec{\ptot}_{\text{cone}})_{\text{T}}$}, where $\vec{\ptot}$ is the momentum of the lepton candidate and $\vec{\ptot}_{\text{cone}}$ is the sum of all the momenta of charged tracks, excluding the lepton candidates, within a distance $\Delta R = \sqrt{\Delta \eta^2 + \Delta \phi ^2}$ of 0.5 around the lepton, where $\Delta \eta$ and $\Delta \phi$ are the pseudorapidity and azimuthal angle differences between the lepton candidate and the charged tracks. The subscript T indicates the momentum component in the transverse plane. A value of $I = 1$ denotes a fully isolated lepton. Candidates with $I(\mu) > 0$ and $I(e) > 0.4$ are selected. Particle identification criteria are applied to the muon and the electron candidates. A tighter identification criterion on the electron is needed to reject the background due to misidentified pions or kaons. This criterion is optimised to preserve signal efficiency while maximising the rejection power over a data sample of same-sign candidates, $\electron^{\pm}\muon^{\pm}$, used as background proxy. The signal selection is also applied on the same-sign candidates.
Figure~\ref{fig:variables} compares distributions of observables for data and simulated \bbbar candidates, and examples of signals with different \mLLP and \tauLLP values, which survive the selection presented above.
Figures~\ref{fig:variables}(a) and (b) show the candidates \mcorr and flight distance distributions. These observables are used in the fit to determine the presence of signal, as explained in Sect.~\ref{sec:signal_yield}. Figures~\ref{fig:variables}(c) and (d) show the transverse momentum distributions of the muon and electron, respectively. These muon and electron \pt distributions show the effect of the \pt threshold in the muon and the electron triggers.
In Figs.~\ref{fig:variables}(e) and (f) the distributions of the isolation variable, $I$, are displayed for the muon and electron, respectively. The leptons from the signal are expected to be more isolated than the ones from the \bbbar background.

\begin{figure}[tb]
  \begin{center}
  
        \includegraphics[width=\textwidth]{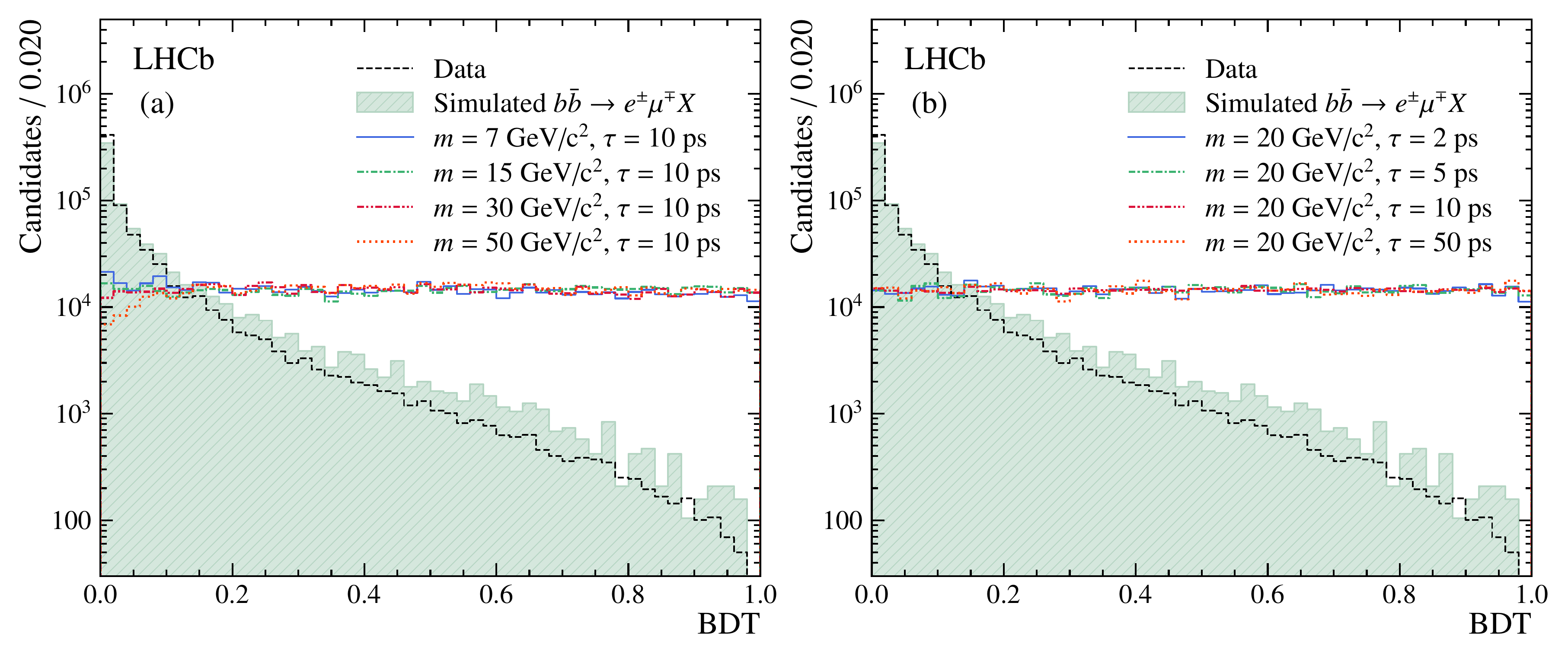}
        
  \vspace*{-0.5cm}
  \end{center}
  \caption{
    \small 
    Distribution of the BDT response in data (dashed black histogram) compared to simulated $\bbbar \to \epm \mump X$ (green filled histogram) and LLP signal samples (coloured histograms) for different (a) \mLLP and (b) \tauLLP values, where the LLP is produced through the DPP mechanism. The distributions from simulation are normalised to the number of candidates in data.
    }
  \label{fig:bdt}
\end{figure}

A Boosted Decision Tree (BDT) classifier \cite{Breiman, AdaBoost} is used to further purify the \mbox{\LLPdec} candidate sample. The BDT is trained using 70k signal decays from a combination of DPP samples, and background candidates drawn from the same-sign sample. The full signal sample contains 2000 candidates for each set of (\mLLP, \tauLLP) parameters. Using all simulated signal samples for the training phase allows to obtain a uniform BDT response across the (\mLLP, \tauLLP) space. Furthermore, the uniformity is enforced by using a special cost function described in Ref.~\cite{Rogozhnikov:2016bdp}. This cost function has the objective to provide the best classification between the signal and the background, while keeping the BDT response uniform on \mLLP and \tauLLP. The BDT input observables are: the muon \pt; the maximum between the momentum of the two leptons; the two isolation variables; the angle between the muon momentum in the $e\mu$ rest frame and the $e\mu$ momentum; the ratio of the energy deposited by the muon in the calorimeters and its momentum; the ratio of the energy deposited by the electron in the \hcal and its momentum; the distance of closest approach between the two lepton tracks; the \chisq of the LLP decay vertex; the difference between the muon and electron impact parameters divided by the LLP impact parameter; the impact parameter \chisq of the leptons, $\chisqip(l)$, divided by $\chisqip(\text{LLP})$. For a given particle, the impact parameter \chisq is defined as the difference between the \chisq of the PV reconstructed with and without that particle.
The BDT response, shown in Fig.~\ref{fig:bdt}, is uniformly distributed between 0 and 1 for the signal, while peaking at zero for the background. Candidates with a BDT value below $0.1$ are rejected, leaving 61116 signal candidates. The observed BDT distribution is consistent with a \bbbar composition of the background. Using the \bbbar cross-section at $13 \tev$ measured by \lhcb, $144 \pm 1 \pm 21 \mub$ \cite{LHCb-PAPER-2016-031}, $(60 \pm 14) \times 10^3$ $\bbbar \to \epm \mump X$ candidates are predicted after selection, consistent with the observed total yield. 

\section{Determination of the signal yield}
\label{sec:signal_yield}

The signal yield is determined from a simultaneous extended maximum likelihood fit to the LLP corrected mass \mcorr and flight distance distributions selected into two BDT intervals (0.1, 0.5] and (0.5, 1.0]. 
The study of the simulated $\bbbar \to \epm \mump X$ background indicates the presence of two components that depend on whether the two leptons belong to the same heavy-flavour jet or two different jets. The two components have different \mcorr and flight distance distributions, and can be separated by the distance $\dr$ between the two leptons. When leptons originate from the same heavy-flavour jet, they have relatively small $\dr$, selected with $\dr < 1$, while $\dr \geq 1$ selects the complementary component.
The background probability density functions of the \mcorr and flight distance needed in the global fit are inferred from the same-sign data. This choice has been validated by a comparison of the distributions of \mcorr and the flight distance in simulated $\bbbar \to \epm \mump X$ and $\bbbar \to \epm \mupm X$ candidates.

When $\dr < 1$, the background \mcorr values are mostly found below $6 \gevcc$. This component is modelled using a sum of a Gaussian and a Crystal Ball function \cite{Skwarnicki:1986xj}. The fraction between the two distributions is fixed to the value obtained in the fit to the same-sign data. The parameters describing the tail are free in each BDT bin. Other parameters are free but common to all the BDT bins.
For the $\dr \geq 1$ region \mcorr is mostly above $10 \gevcc$. This region is modelled using a Johnson S$_{U}$ distribution \cite{10.2307/2332669} with shape parameters free in each BDT bin.
To model the signal \mcorr distribution a sum of a modified Gaussian distribution, where the left tail is exponential and the right tail a power law, and another Gaussian distribution is used. The parameters of the model are fixed to the values obtained from the fits to the simulated samples, for each (\mLLP, \tauLLP) hypothesis. The same signal \mcorr models are used for each BDT bin and production mechanism.

The background candidates with $\dr < 1$  have long flight distances, above $10 \mm$. The opposite is true for $\dr \geq 1$. The two components are modelled using a \mbox{Johnson S$_{U}$} distribution, with all parameters kept free. In the $\dr < 1$ region the parameters of the model are not shared across the BDT bins, while they are shared when $\dr \geq 1$.
A kernel density estimation algorithm is used to estimate the probability density function of the flight distance distribution in simulated signal for each BDT bin. The same signal flight distance model for a given (\mLLP, \tauLLP) hypothesis is used for each production mechanism. 

\begin{figure}[tb]
   \begin{center}
  
         \includegraphics[width=\textwidth]{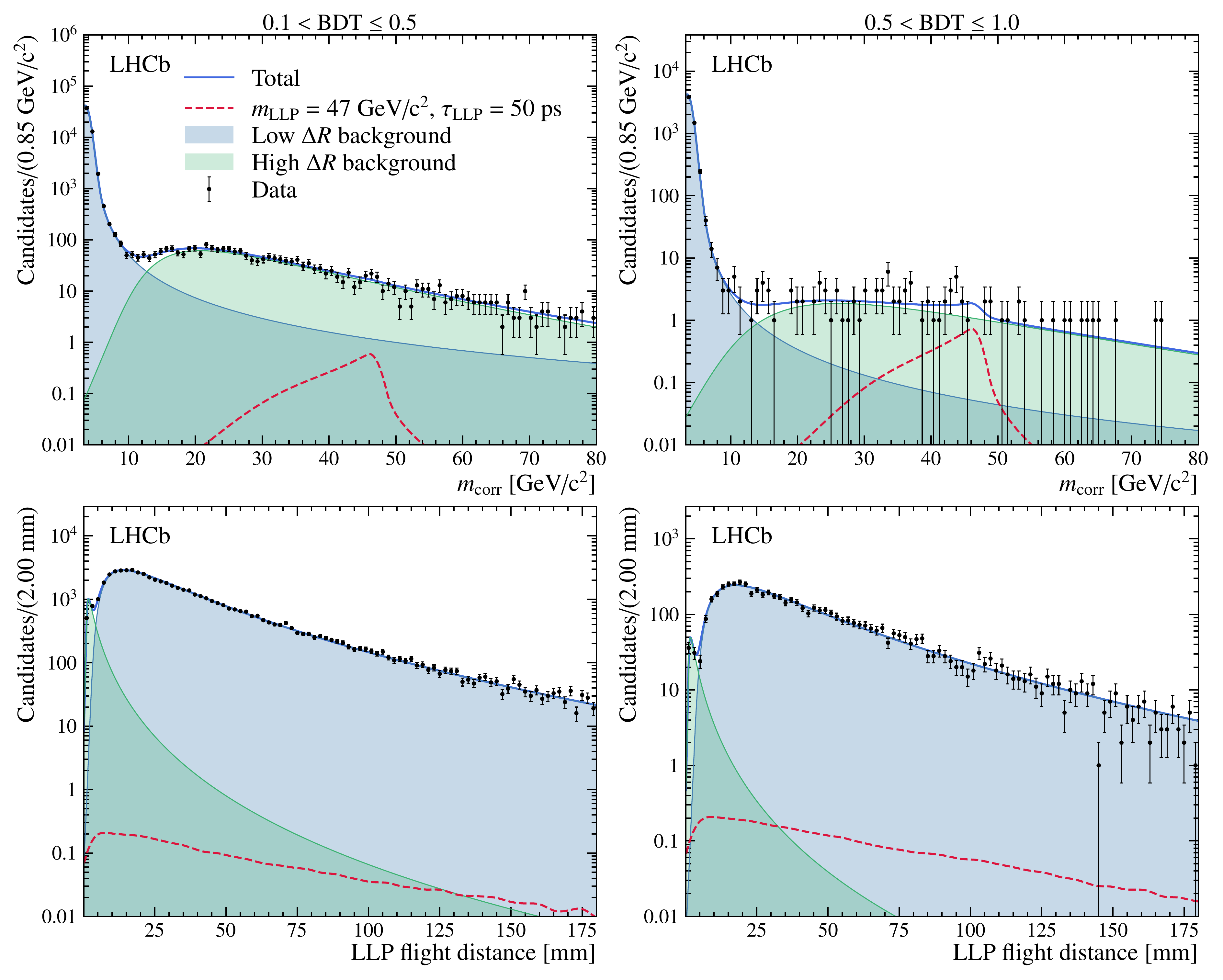}
        
   \vspace*{-0.5cm}
   \end{center}
   \caption{
     \small
     Distributions of \mcorr (top) and the flight distance (bottom) of two BDT intervals (left and right), where a  simultaneous fit result for a LLP signal with $\mLLP = 47 \gevcc$ and $\tauLLP = 50 \ps$ is overlaid; the fitted signal yield in this example is $14 \pm 14$.
     }
   \label{fig:fit_example}
 \end{figure}

In the final fit the fractions of signal yield in each BDT interval are constrained by Gaussian functions to the values and uncertainties that are estimated in the simulation. In order to explore a larger set of \mLLP values than the simulated set, signal templates for the \mcorr and flight distance distributions are interpolated from the simulated distributions using a moment morphing algorithm \cite{BAAK201539}. Distributions of \mcorr and the flight distance in two BDT regions are shown in Fig.~\ref{fig:fit_example}, with an example of a fit result for a signal with $\mLLP = 47 \gevcc$ and $\tauLLP = 50 \ps$ overlaid. For each \mLLP and \tauLLP hypothesis the fitted yields are consistent with no signal present.

\section{Signal efficiencies and systematic uncertainties}

The determination of the signal detection efficiency relies on simulation.
Systematic effects are identified from differences between data and simulation. Regarding the electron, samples of $\jpsi \to \epem$ and \Zee decays are considered, and $\jpsi \to \mumu$, \mbox{\Upsmumu} and \Zmumu decays are used for the muon. Samples of $\bbbar \to \epm \mupm X $ candidates are used to compare distributions of the reconstructed dilepton system such as the corrected mass and the flight distance.
Systematic uncertainties on the signal efficiency have been evaluated. They are summarised in Table~\ref{tab:systematics} and discussed in more details below. Also reported in the table are the uncertainties on the integrated luminosity, evaluated to be $2\%$ \cite{LHCB-PAPER-2014-047}, on the signal fraction in each BDT bin, and on the signal yield associated with the fit procedure, discussed at the end of this section. 

To account for the mismodelling in the  simulation used to compute the signal efficiency, a bias for each variable used in the selection is determined by comparing simulated and experimental distributions of \Z and \bbbar candidates. The correlations between the selection variables are computed using the signal samples. The effect of imperfect simulation is subsequently estimated by recomputing several times the signal efficiency after changing the selection requirements on the variables by factors drawn from a multivariate normal distribution, with biases and correlations between the variables as input. The standard deviation of the distribution of efficiencies is found in the range $4.9$ to $7.3\%$, depending on the signal mass, lifetime and production mechanism, which is taken as a contribution to the systematic uncertainty. In a similar way, systematic uncertainties ranging from $0.5$ to $2.4\%$ are assigned to the identification of the two leptons.

The systematic uncertainty due to the imprecision in the simulated signal sample used to train the BDT classifier is estimated by applying the classifier on modified signal distributions: each input variable is multiplied by a scale factor drawn from a multivariate normal distribution built with the variable biases and correlations, also inferred from the control samples. The standard deviation of the efficiency distribution is used as systematic uncertainty, ranging from $0.6$ to $1.0\%$ for the BDT $> 0.1$ requirement, and from $3.3$ to $4.0\%$ on the signal fraction in the BDT bins.

\noindent The contribution to the systematic uncertainty from the statistical precision of the simulated signal samples is in the range \numrange{1.1}{3.0}$\%$.

The theoretical uncertainties are dominated by the limited knowledge of the partonic luminosity. This contribution is estimated following the procedure explained in Ref.~\cite{Butterworth:2015oua} and varies from $1.1\%$ up to $6.1\%$. The minimum systematic contribution is found for the DPP and CC processes while the maximum contribution is found for the gluon-gluon fusion process HIG.

\begin{figure}[tb]
   \begin{center}
  
         \includegraphics[width=\textwidth]{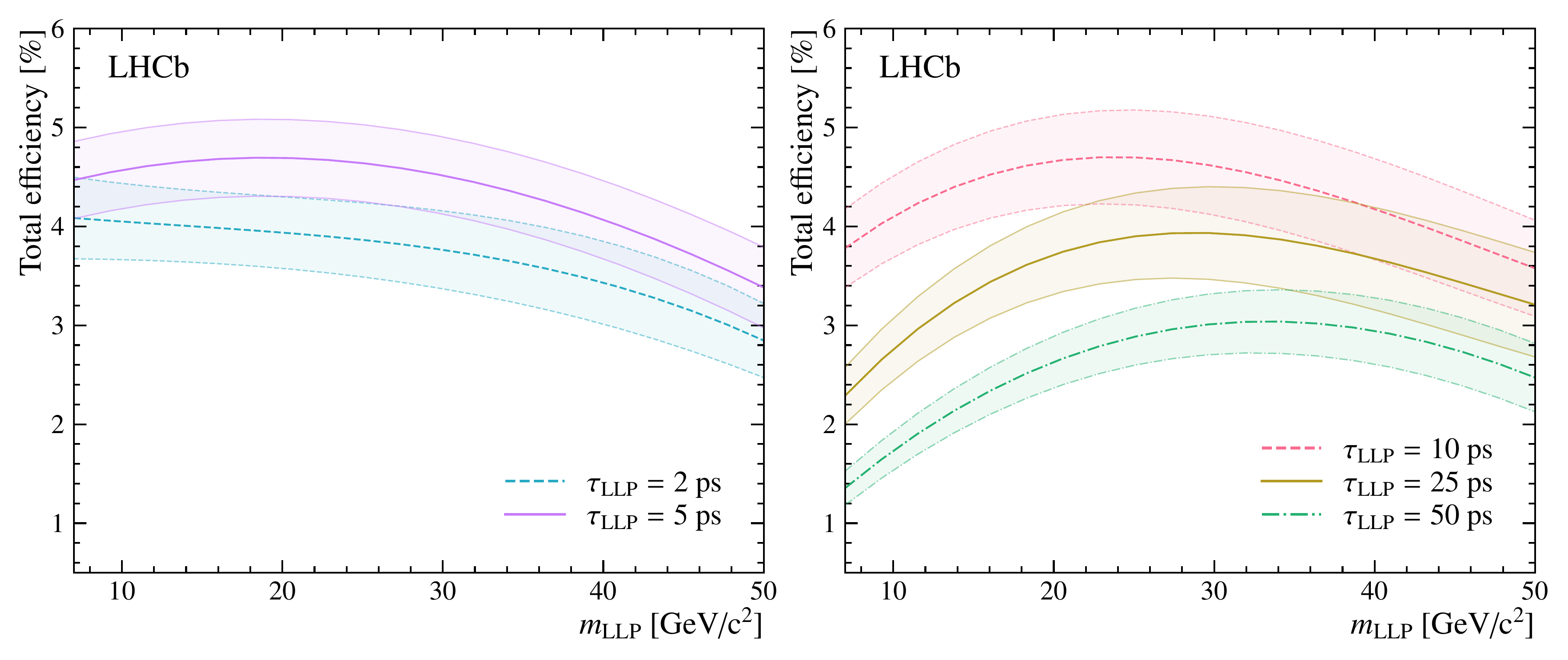}
        
   \vspace*{-0.5cm}
   \end{center}
   \caption{
     \small
     Total detection efficiency for LLP produced through the DPP mechanism as a function of \mLLP (central line) and its uncertainty (coloured band), obtained for different values of \tauLLP.
     }
   \label{fig:dpp_efficiencies}
 \end{figure}

Finally, the total systematic uncertainty is obtained as the sum in quadrature of all contributions, where the different components of the detection efficiency are assumed to be fully correlated. In order to uniformly cover the full \mLLP range, a third-order polynomial is fitted to the signal detection efficiency as function of \mLLP for each simulated \tauLLP value. A second order polynomial is also fitted to the efficiency. The difference between the two efficiencies is assigned as systematic uncertainty, a contribution that is always less than $4 \%$.
The interpolated signal efficiency for LLPs produced through the DPP mechanism is shown in Fig.~\ref{fig:dpp_efficiencies}, accounting for the geometrical acceptance. The criteria on the vertex displacement favour large lifetimes; however, above $10 \ps$ the probability that the LLP decays outside the \velo increases, leading to a loss of efficiency. The selection efficiency increases with \mLLP, however, this effect is counteracted by the loss of lepton candidates outside the spectrometer acceptance, which is more likely for heavier LLPs. Therefore the signal efficiencies are highest for masses between 20 and $30 \gevcc$ and lifetimes between 5 and $10 \ps$. The DPP mechanism has the highest detection efficiency. On average, the detection efficiency for the HIG (CC) mechanism is $20\%$ ($60\%$) lower than the DPP mechanism.

\begin{table}[tb]
\begin{center}
\caption{\small Contributions to the relative systematic uncertainties in $\%$. The contributions are grouped in three categories, the integrated luminosity, the detection efficiency and the signal yield, separated by horizontal lines. The detection efficiency is affected by the parton luminosity model and depends upon the production process, with a maximum uncertainty of $6.1\%$ for the gluon-gluon fusion process HIG.}
\label{tab:systematics}
\begin{tabular}{cc}
\toprule
Source & Contribution [$\%$] \\ 
\midrule
Integrated luminosity & $2.0$ \\ 
\midrule
Reconstruction and selection & \numrange{4.9}{7.3} \\ 
Particle identification & \numrange{0.5}{2.4} \\ 
BDT & \numrange{0.6}{1.0} \\ 
Simulation sample size & \numrange{1.1}{3.0} \\
Parton luminosity & \numrange{1.1}{6.1} \\ 
Efficiency interpolation & \numrange{0.1}{4.0} \\
\midrule
Signal fraction in the BDT bins & \numrange{3.3}{4.0} \\
Signal model & \numrange{0.7}{8.1} \\
\midrule
Total & \numrange{10.6}{17.7} \\ 
\bottomrule
\end{tabular}
\end{center}
\end{table}

The choice of templates for the corrected mass and flight distance can affect the result of the fit. The uncertainty due to the signal model accounts for imperfect simulation of the scale and resolution of the \mcorr and flight distance, and that of the finite size of the simulated signal samples used to produce the probability density functions.
Uncertainties of $0.2\%$ on the \mcorr scale and $1.6\%$ on the \mcorr resolution are estimated from the comparison between data and \bbbar simulated candidates. For the flight distance a scale uncertainty of $1.2 \%$ and a resolution uncertainty of $1.1 \%$ are estimated.  
The propagation of uncertainties is performed using pseudoexperiments generated from the background model fitted to the same-sign data. Ten signal data points are drawn from modified signal \mcorr and flight distance distributions, modified by smearing or rescaling, and added to each pseudoexperiment. The fitted signal yield is compared to the result with ten signal data points drawn from a non-modified signal. Changing the \mcorr scale leads to a relative change on the signal yield from $0.1$ to $1.2\%$, and $0.1$ to $0.8\%$ for the flight distance, depending on the signal hypothesis. A relative variation of the signal yield from $0.1$ to $8.1\%$ is observed from an additional smearing of the signal \mcorr distribution, $0.1$ to $0.8\%$ for the flight distance. 
The effect of the limited sample size used to construct the signal model is addressed by replacing the parameter values of the signal model by values drawn from Gaussian distributions. For each parameter the mean of the Gaussian distribution is equal to its fitted value, and the standard deviation is equal to its uncertainty. A relative variation of the signal yield due to the limited sample size is found to be between $0.1$ and $1.7\%$. A total systematic uncertainty \numrange{0.7}{8.1}$\%$ is accounted for the signal yield.

All the systematic uncertainties related to the integrated luminosity, the signal efficiency and the signal yield are included as nuisance parameters in the determination of the cross-section upper limits.

\section{Results}

\begin{figure}[tb]
   \begin{center}
  
         \includegraphics[width=\textwidth]{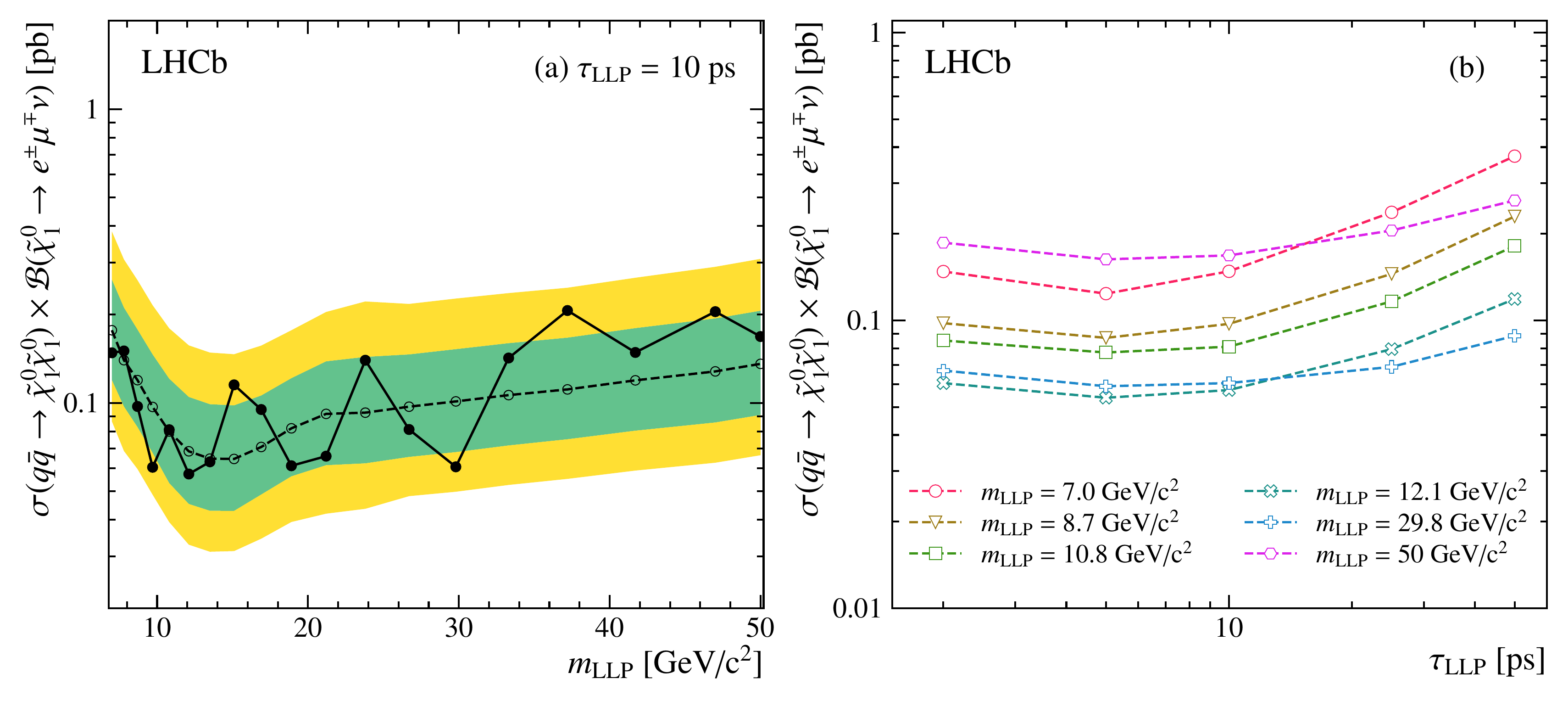}
        
   \vspace*{-0.5cm}
   \end{center}
   \caption{
     \small 
     (a) Expected (open circles and dotted line) and observed (filled circles and solid line) upper limits of the cross-section as a function of \mLLP for $\tauLLP = 10 \ps$, for LLPs produced through the DPP mechanism. The green and yellow bands indicate the quantiles of the expected upper limit corresponding to $\pm1\sigma$ and $\pm2\sigma$ for a Gaussian distribution. (b) Observed limits on the cross-section as a function of \tauLLP for different \mLLP values for LLPs produced through the DPP mechanism.
     }
   \label{fig:limits}
 \end{figure}

\begin{figure}[tb]
   \begin{center}
  
         \includegraphics[width=\textwidth]{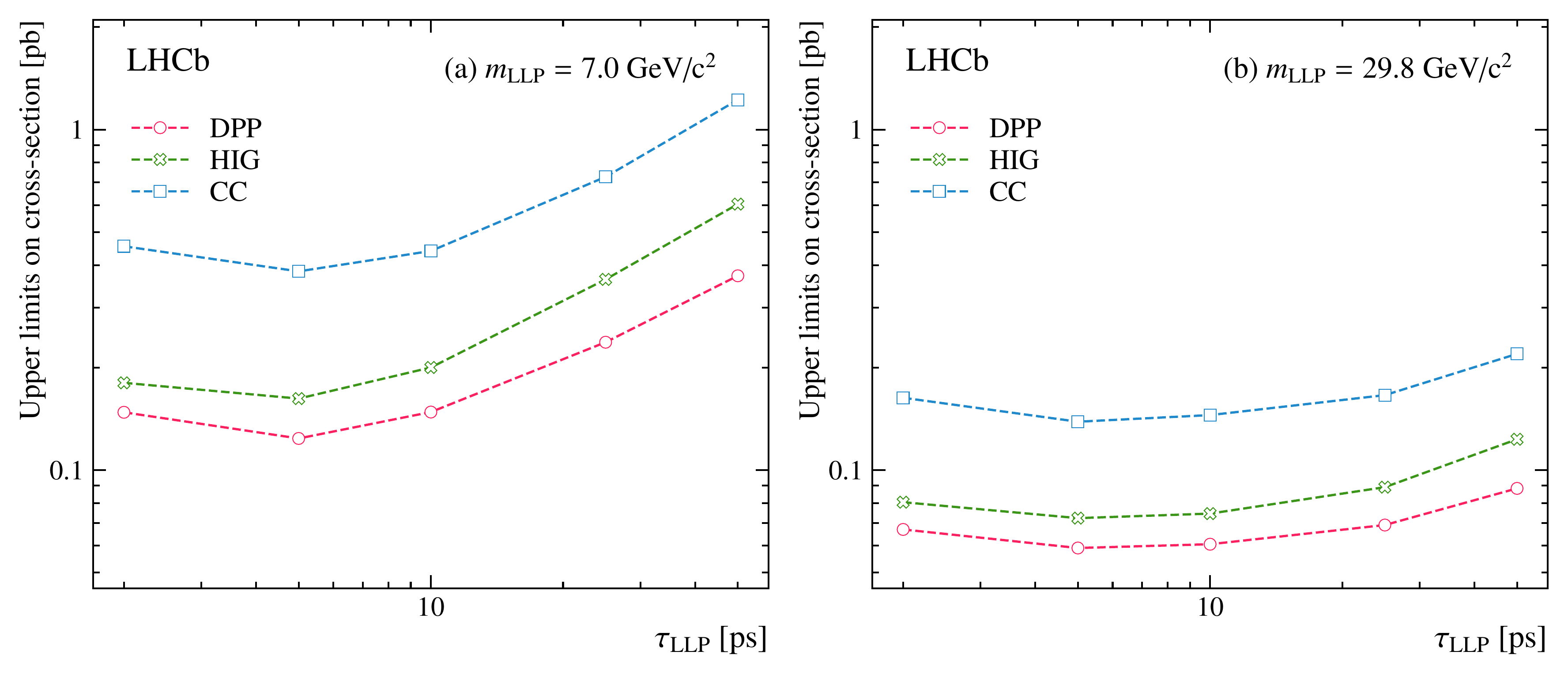}
        
   \vspace*{-0.5cm}
   \end{center}
   \caption{
     \small 
     Observed upper limits on the production cross-sections times branching fraction for (a) $\mLLP = 7 \gevcc$ and (b) $\mLLP = 29.8 \gevcc$ as function of \tauLLP for the DPP, HIG and CC production mechanisms.
     }
   \label{fig:limits_prod}
 \end{figure}

\begin{figure}[tb]
   \begin{center}
  
         \includegraphics[width=0.55\textwidth]{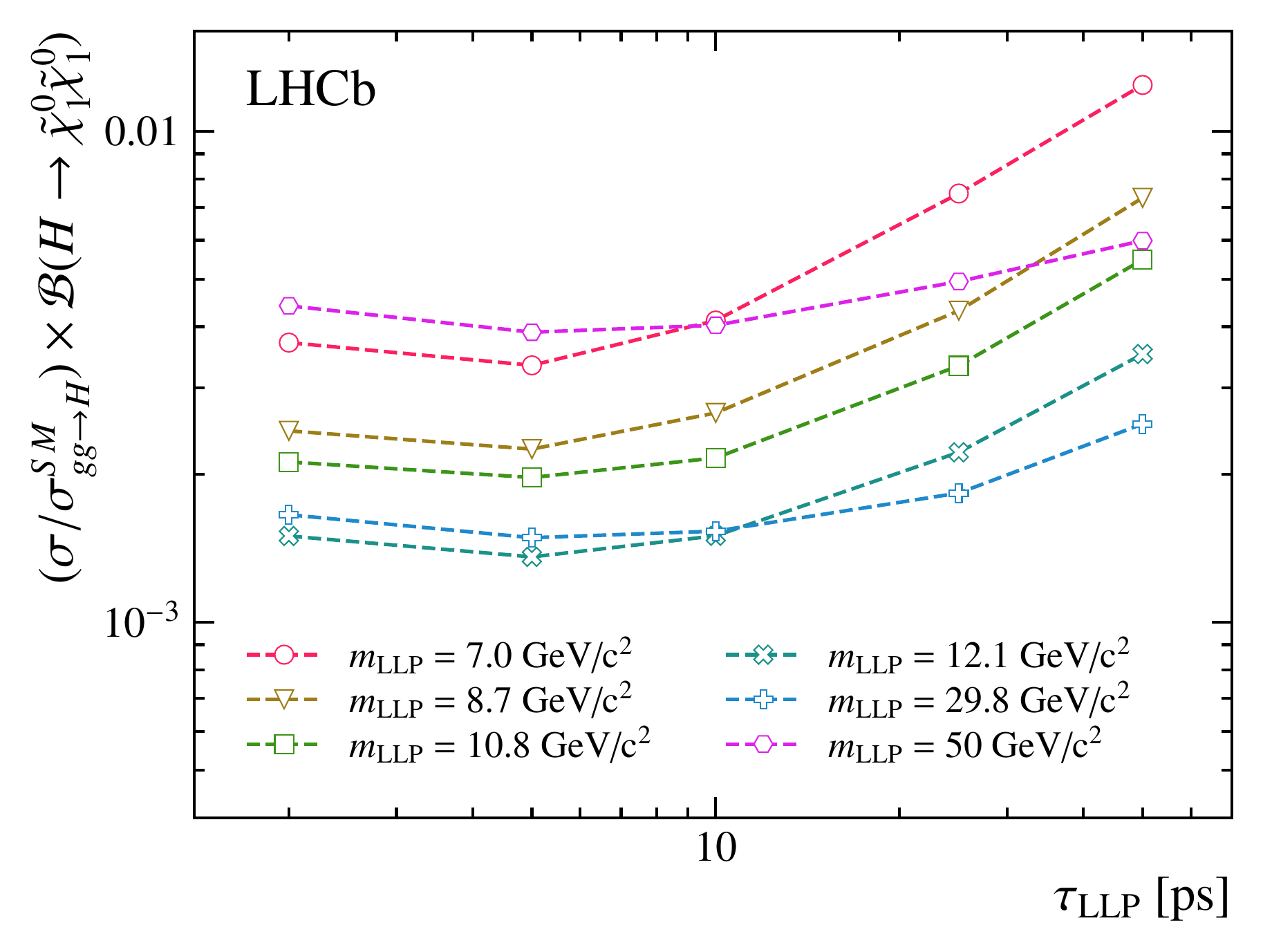}
        
   \vspace*{-0.5cm}
   \end{center}
   \caption{
     \small 
     Observed limits on the $(\sigma / \sigma^{SM}_{gg \rightarrow H}) \times \mathcal{B}(\H \rightarrow \tilde{\chi}^{0}_{1} \tilde{\chi}^{0}_{1})$, assuming $\mathcal{B}(\tilde{\chi}^{0}_{1} \rightarrow e^{\pm}\mu^{\mp}\nu) = 1$ as a function of \tauLLP for different \mLLP values. The value of the gluon-gluon fusion production cross-section used is $48.6 \pm 3.5 \pb$ \cite{Cepeda:2019klc}.
     }
   \label{fig:limits_higgs_br}
 \end{figure}

The results of the simultaneous fits to the LLP corrected mass and flight distance distributions in the two BDT intervals (0.1, 0.5] and (0.5, 1.0], are found to be compatible with the background-only hypothesis for all signal hypotheses considered. Upper limits at 95$\%$ Confidence Level (CL) on the production cross-sections times branching fraction are computed for each production mechanism,
\begin{align*}
    \sigma_{\text{DPP}} &= \sigma(q\bar{q} \rightarrow \tilde{\chi}^{0}_{1}\tilde{\chi}^{0}_{1}) \times \mathcal{B}(\tilde{\chi}^{0}_{1} \rightarrow e^{\pm}\mu^{\mp}\nu), \\
    \sigma_{\text{HIG}} &= \sigma(gg \rightarrow h) \times \mathcal{B}(h \rightarrow \tilde{\chi}^{0}_{1} \tilde{\chi}^{0}_{1}) \times \mathcal{B}(\tilde{\chi}^{0}_{1} \rightarrow e^{\pm}\mu^{\mp}\nu), \text{ and}\\
    \sigma_{\text{CC}} &= \sigma(\W \to l N) \times \mathcal{B}(N \rightarrow e^{\pm}\mu^{\mp}\nu),
\end{align*}

\noindent for each pair of \mLLP and \tauLLP values using the CLs approach \cite{CLs}. Upper limits for selected \mLLP and \tauLLP values are shown in Figs.~\ref{fig:limits} to \ref{fig:limits_higgs_br}.
Figure~\ref{fig:limits}(a) gives examples of observed upper limits on $\sigma_{\text{DPP}}$, along with the range of limits expected for the background-only hypothesis, as a function of \mLLP for $\tauLLP = 10 \ps$. Figure~\ref{fig:limits}(b) shows the observed upper limits on $\sigma_{\text{DPP}}$ as a function of \tauLLP, for a selection of \mLLP values that shows the range of  limit values. The best observed limits on $\sigma_{\text{DPP}}$ are of the order of $0.06 \pb$ for a mass of $29.8 \gevcc$. A comparison of observed upper limits on $\sigma_{\text{DPP}}$, $\sigma_{\text{HIG}}$ and $\sigma_{\text{CC}}$ as a function of \tauLLP for the lowest mass studied, $\mLLP = 7$, and $29.8 \gevcc$ is shown in Fig.~\ref{fig:limits_prod}. The best and worst limits are obtained for the DPP and CC mechanisms, respectively. The differences between the sensitivities for each production mechanism are principally due to detection efficiency. The limits obtained by the \atlas experiment on the squark-antisquark production cross-section \cite{Aad:2019tcc}, where the squark has a mass of 700 or $1600\gevcc$ and decays to $\quark \, (\tilde{\chi}^{0}_{1} \rightarrow ee\nu / e\mu\nu / \mu\mu\nu)$, have values from $1$ to $10 \fb$ for $m(\tilde{\chi}^{0}_{1}) = 50 \gevcc$ in the lifetime range studied. These results are complementary to the results obtained by the \atlas experiment, extend to lower mass and lifetime regions and explore different LLP production mechanisms.

Finally, the limits on $\sigma_{\text{HIG}}$ are compared to the value of the SM Higgs boson production cross-section from gluon-gluon fusion of $48.6 \pm 3.5 \pb$ \cite{Cepeda:2019klc}, which is illustrated in Fig.~\ref{fig:limits_higgs_br}. These limits are placed on $(\sigma / \sigma^{SM}_{gg \rightarrow H}) \times \mathcal{B}(\H \rightarrow \tilde{\chi}^{0}_{1} \tilde{\chi}^{0}_{1})$, assuming $\mathcal{B}(\tilde{\chi}^{0}_{1} \rightarrow e^{\pm}\mu^{\mp}\nu) = 1$, as a function of \tauLLP for a selection of \mLLP values. Under this assumption the limits on $\mathcal{B}(\H \rightarrow \tilde{\chi}^{0}_{1} \tilde{\chi}^{0}_{1})$ have a minimum of $\sim 0.15 \%$. Decays of $\text{LLP} \to \mumu$, produced in pairs from SM Higgs bosons, were searched by the \cms experiment \cite{CMS:2014hka}. Assuming $\mathcal{B}(\text{LLP} \rightarrow \mumu) = 1$, the limits on $\mathcal{B}(\H \rightarrow \text{LLP} \, \text{LLP})$ for $\mLLP = 50 \gevcc$ are the best for lifetimes between $1\ps$ and $10 \ns$ with a minimum of $0.05 \%$ \cite{Lee:2018pag}, which is approximately 3 times lower than the minimum limits on $\mathcal{B}(\H \rightarrow \tilde{\chi}^{0}_{1} \tilde{\chi}^{0}_{1})$ presented in this paper.

\section{Conclusion}
A search for decays of long-lived massive particles, in the $\epm \mump \neu$ final state, is performed using $\proton\proton$ collisions at $\sqs =13 \tev$ recorded with the LHCb detector, for a total integrated luminosity of $5.38 \pm 0.11 \invfb$. The search covers LLP masses from 7 to $50 \gevcc$, lifetimes from 2 to $50 \ps$ and considers three production mechanisms: the direct pair production from the interaction of quarks, the pair production from the decay of a SM-like Higgs boson with a mass of $125 \gevcc$, and the charged current production from an on-shell \W boson with an additional lepton.

Fully simulated signal events are used to define the signal selection criteria and the signal detection efficiency. The background is dominated by $\bbbar$ candidates. A BDT, taking as input properties of the leptons and displaced vertex of the LLP, is used to purify the signal from the heavy hadron background. The signal yield is determined by a simultaneous fit of the LLP corrected mass and flight distance, using signal templates derived from simulation. All the results of the fits are compatible with the absence of signal, and upper limits on the cross-section times branching fraction for each production mechanism are computed. The best upper limits are achieved for the pair production, from interaction of quarks or the decay of a SM-like Higgs boson, for lifetimes below $10 \ps$ and masses above $10 \gevcc$, and are of the order of $0.1 \pb$.

\section*{Acknowledgements}
%
%
\noindent We express our gratitude to our colleagues in the CERN
accelerator departments for the excellent performance of the LHC. We
thank the technical and administrative staff at the LHCb
institutes.
We acknowledge support from CERN and from the national agencies:
CAPES, CNPq, FAPERJ and FINEP (Brazil); 
MOST and NSFC (China); 
CNRS/IN2P3 (France); 
BMBF, DFG and MPG (Germany); 
INFN (Italy); 
NWO (Netherlands); 
MNiSW and NCN (Poland); 
MEN/IFA (Romania); 
MSHE (Russia); 
MICINN (Spain); 
SNSF and SER (Switzerland); 
NASU (Ukraine); 
STFC (United Kingdom); 
DOE NP and NSF (USA).
We acknowledge the computing resources that are provided by CERN, IN2P3
(France), KIT and DESY (Germany), INFN (Italy), SURF (Netherlands),
PIC (Spain), GridPP (United Kingdom), RRCKI and Yandex
LLC (Russia), CSCS (Switzerland), IFIN-HH (Romania), CBPF (Brazil),
PL-GRID (Poland) and OSC (USA).
We are indebted to the communities behind the multiple open-source
software packages on which we depend.
Individual groups or members have received support from
AvH Foundation (Germany);
EPLANET, Marie Sk\l{}odowska-Curie Actions and ERC (European Union);
A*MIDEX, ANR, Labex P2IO and OCEVU, and R\'{e}gion Auvergne-Rh\^{o}ne-Alpes (France);
Key Research Program of Frontier Sciences of CAS, CAS PIFI,
Thousand Talents Program, and Sci. \& Tech. Program of Guangzhou (China);
RFBR, RSF and Yandex LLC (Russia);
GVA, XuntaGal and GENCAT (Spain);
the Royal Society
and the Leverhulme Trust (United Kingdom).

\addcontentsline{toc}{section}{References}
\bibliographystyle{LHCb}
\bibliography{main,standard,LHCb-PAPER,LHCb-CONF,LHCb-DP,LHCb-TDR}

\newpage
\centerline
{\large\bf LHCb collaboration}
\begin
{flushleft}
\small
R.~Aaij$^{31}$,
C.~Abell{\'a}n~Beteta$^{49}$,
T.~Ackernley$^{59}$,
B.~Adeva$^{45}$,
M.~Adinolfi$^{53}$,
H.~Afsharnia$^{9}$,
C.A.~Aidala$^{84}$,
S.~Aiola$^{25}$,
Z.~Ajaltouni$^{9}$,
S.~Akar$^{64}$
% ...

{\footnotesize \it

% ...
$ ^{9}$Universit{\'e} Clermont Auvergne, CNRS/IN2P3, LPC, Clermont-Ferrand, France\\
% ...
$ ^{25}$INFN Sezione di Milano, Milano, Italy\\
% ...
$ ^{31}$Nikhef National Institute for Subatomic Physics, Amsterdam, Netherlands\\
% ...
$ ^{45}$Instituto Galego de F{\'\i}sica de Altas Enerx{\'\i}as (IGFAE), Universidade de Santiago de Compostela, Santiago de Compostela, Spain\\
% ...
$ ^{49}$Physik-Institut, Universit{\"a}t Z{\"u}rich, Z{\"u}rich, Switzerland\\
% ...
$ ^{53}$H.H. Wills Physics Laboratory, University of Bristol, Bristol, United Kingdom\\
% ...
$ ^{59}$Oliver Lodge Laboratory, University of Liverpool, Liverpool, United Kingdom\\
% ...
$ ^{64}$University of Cincinnati, Cincinnati, OH, United States\\
% ...
$ ^{67}$Syracuse University, Syracuse, NY, United States\\
% ...
$ ^{84}$University of Michigan, Ann Arbor, United States, associated to $^{67}$\\
\bigskip
% ...
\medskip
}
\end{flushleft}




\end{document}


\def\mLLP {\ensuremath{m_{\text{LLP}}}\xspace}
\def\tauLLP {\ensuremath{\tau_{\text{LLP}}}\xspace}

\section*{Supplementary material for LHCb-PAPER-2020-027}

Tables~\ref{tab:sup_dpp_eff_limit}, \ref{tab:sup_hig_eff_limit} and \ref{tab:sup_cc_eff_limit} present the detection efficiencies and the upper limits on production cross-section times branching fraction as a function of \mLLP and \tauLLP for the LLPs produced through the DPP, HIG and CC mechanisms, respectively.
\begin{table}
\caption{Efficiency in percentage, and 95 $\%$ CL upper limits on the production cross-section times branching fraction for LLPs produced through the DPP mechanism.}
\renewcommand{\arraystretch}{1.12}
\begin{minipage}{.48\linewidth}
\resizebox{\textwidth}{!}{
\begin{tabular}{cccccc}
\toprule
\mLLP & \tauLLP & $\epsilon$ & expected UL & observed UL \\ 
\midrule
7.0 & 2 & $4.1 \pm 0.4$ & $0.16^{+0.08}_{-0.05}$ & 0.15 \\ 
7.0 & 5 & $4.5 \pm 0.4$ & $0.14^{+0.07}_{-0.04}$ & 0.12 \\ 
7.0 & 10 & $3.8 \pm 0.4$ & $0.18^{+0.09}_{-0.06}$ & 0.15 \\ 
7.0 & 25 & $2.3 \pm 0.3$ & $0.27^{+0.14}_{-0.09}$ & 0.24 \\ 
7.0 & 50 & $1.4 \pm 0.2$ & $0.47^{+0.24}_{-0.15}$ & 0.37 \\ 
7.8 & 2 & $4.1 \pm 0.4$ & $0.13^{+0.07}_{-0.04}$ & 0.16 \\ 
7.8 & 5 & $4.5 \pm 0.4$ & $0.12^{+0.06}_{-0.04}$ & 0.13 \\ 
7.8 & 10 & $3.9 \pm 0.4$ & $0.14^{+0.07}_{-0.04}$ & 0.15 \\ 
7.8 & 25 & $2.4 \pm 0.3$ & $0.23^{+0.12}_{-0.07}$ & 0.23 \\ 
7.8 & 50 & $1.5 \pm 0.2$ & $0.37^{+0.18}_{-0.12}$ & 0.35 \\ 
8.7 & 2 & $4.1 \pm 0.4$ & $0.11^{+0.06}_{-0.04}$ & 0.10 \\ 
8.7 & 5 & $4.5 \pm 0.4$ & $0.10^{+0.05}_{-0.03}$ & 0.09 \\ 
8.7 & 10 & $4.0 \pm 0.4$ & $0.12^{+0.06}_{-0.04}$ & 0.10 \\ 
8.7 & 25 & $2.6 \pm 0.3$ & $0.19^{+0.09}_{-0.06}$ & 0.15 \\ 
8.7 & 50 & $1.6 \pm 0.2$ & $0.30^{+0.15}_{-0.09}$ & 0.23 \\ 
9.7 & 2 & $4.1 \pm 0.4$ & $0.10^{+0.05}_{-0.03}$ & 0.06 \\ 
9.7 & 5 & $4.6 \pm 0.4$ & $0.09^{+0.04}_{-0.03}$ & 0.06 \\ 
9.7 & 10 & $4.1 \pm 0.4$ & $0.10^{+0.05}_{-0.03}$ & 0.06 \\ 
9.7 & 25 & $2.7 \pm 0.3$ & $0.15^{+0.08}_{-0.04}$ & 0.09 \\ 
9.7 & 50 & $1.7 \pm 0.2$ & $0.24^{+0.12}_{-0.07}$ & 0.14 \\ 
10.8 & 2 & $4.0 \pm 0.4$ & $0.08^{+0.04}_{-0.03}$ & 0.09 \\ 
10.8 & 5 & $4.6 \pm 0.4$ & $0.07^{+0.04}_{-0.02}$ & 0.08 \\ 
10.8 & 10 & $4.2 \pm 0.4$ & $0.08^{+0.04}_{-0.03}$ & 0.08 \\ 
10.8 & 25 & $2.9 \pm 0.3$ & $0.12^{+0.06}_{-0.04}$ & 0.12 \\ 
10.8 & 50 & $1.8 \pm 0.2$ & $0.19^{+0.10}_{-0.06}$ & 0.18 \\ 
12.1 & 2 & $4.0 \pm 0.4$ & $0.07^{+0.04}_{-0.02}$ & 0.06 \\ 
12.1 & 5 & $4.6 \pm 0.4$ & $0.06^{+0.03}_{-0.02}$ & 0.05 \\ 
12.1 & 10 & $4.3 \pm 0.4$ & $0.07^{+0.04}_{-0.02}$ & 0.06 \\ 
12.1 & 25 & $3.0 \pm 0.3$ & $0.10^{+0.05}_{-0.03}$ & 0.08 \\ 
12.1 & 50 & $2.0 \pm 0.2$ & $0.15^{+0.08}_{-0.05}$ & 0.12 \\ 
13.5 & 2 & $4.0 \pm 0.4$ & $0.07^{+0.04}_{-0.02}$ & 0.07 \\ 
13.5 & 5 & $4.6 \pm 0.4$ & $0.06^{+0.03}_{-0.02}$ & 0.06 \\ 
13.5 & 10 & $4.4 \pm 0.4$ & $0.06^{+0.03}_{-0.02}$ & 0.06 \\ 
13.5 & 25 & $3.2 \pm 0.3$ & $0.09^{+0.05}_{-0.03}$ & 0.09 \\ 
13.5 & 50 & $2.1 \pm 0.2$ & $0.14^{+0.07}_{-0.05}$ & 0.13 \\ 
15.1 & 2 & $4.0 \pm 0.4$ & $0.07^{+0.04}_{-0.02}$ & 0.13 \\ 
15.1 & 5 & $4.7 \pm 0.4$ & $0.06^{+0.03}_{-0.02}$ & 0.11 \\ 
15.1 & 10 & $4.5 \pm 0.4$ & $0.06^{+0.03}_{-0.02}$ & 0.12 \\ 
15.1 & 25 & $3.4 \pm 0.4$ & $0.09^{+0.05}_{-0.03}$ & 0.15 \\ 
15.1 & 50 & $2.3 \pm 0.2$ & $0.13^{+0.07}_{-0.04}$ & 0.23 \\ 
16.9 & 2 & $4.0 \pm 0.4$ & $0.08^{+0.04}_{-0.03}$ & 0.11 \\ 
16.9 & 5 & $4.7 \pm 0.4$ & $0.07^{+0.03}_{-0.02}$ & 0.09 \\ 
16.9 & 10 & $4.6 \pm 0.4$ & $0.07^{+0.04}_{-0.02}$ & 0.09 \\ 
16.9 & 25 & $3.5 \pm 0.4$ & $0.09^{+0.05}_{-0.03}$ & 0.12 \\ 
16.9 & 50 & $2.4 \pm 0.2$ & $0.14^{+0.07}_{-0.04}$ & 0.18 \\ 
18.9 & 2 & $4.0 \pm 0.4$ & $0.10^{+0.05}_{-0.03}$ & 0.07 \\ 
18.9 & 5 & $4.7 \pm 0.4$ & $0.08^{+0.04}_{-0.02}$ & 0.06 \\ 
18.9 & 10 & $4.6 \pm 0.5$ & $0.08^{+0.04}_{-0.03}$ & 0.06 \\ 
18.9 & 25 & $3.6 \pm 0.4$ & $0.10^{+0.05}_{-0.03}$ & 0.08 \\ 
18.9 & 50 & $2.6 \pm 0.3$ & $0.15^{+0.07}_{-0.05}$ & 0.10 \\ 
\bottomrule
\end{tabular}}
\end{minipage}
\hfill
\begin{minipage}{.48\linewidth}
\resizebox{\textwidth}{!}{
\begin{tabular}{cccccc}
\toprule
\mLLP & \tauLLP & $\epsilon$ & expected UL & observed UL \\ 
\midrule
21.2 & 2 & $3.9 \pm 0.4$ & $0.11^{+0.06}_{-0.04}$ & 0.07 \\ 
21.2 & 5 & $4.7 \pm 0.4$ & $0.09^{+0.05}_{-0.03}$ & 0.06 \\ 
21.2 & 10 & $4.7 \pm 0.5$ & $0.09^{+0.05}_{-0.03}$ & 0.07 \\ 
21.2 & 25 & $3.8 \pm 0.4$ & $0.11^{+0.06}_{-0.04}$ & 0.08 \\ 
21.2 & 50 & $2.7 \pm 0.3$ & $0.16^{+0.08}_{-0.05}$ & 0.11 \\ 
23.8 & 2 & $3.9 \pm 0.4$ & $0.11^{+0.06}_{-0.04}$ & 0.15 \\ 
23.8 & 5 & $4.7 \pm 0.4$ & $0.09^{+0.05}_{-0.03}$ & 0.13 \\ 
23.8 & 10 & $4.7 \pm 0.5$ & $0.09^{+0.05}_{-0.03}$ & 0.14 \\ 
23.8 & 25 & $3.9 \pm 0.4$ & $0.11^{+0.06}_{-0.04}$ & 0.16 \\ 
23.8 & 50 & $2.8 \pm 0.3$ & $0.15^{+0.08}_{-0.05}$ & 0.21 \\ 
26.7 & 2 & $3.8 \pm 0.4$ & $0.12^{+0.06}_{-0.04}$ & 0.09 \\ 
26.7 & 5 & $4.6 \pm 0.4$ & $0.10^{+0.05}_{-0.03}$ & 0.08 \\ 
26.7 & 10 & $4.7 \pm 0.5$ & $0.10^{+0.05}_{-0.03}$ & 0.08 \\ 
26.7 & 25 & $3.9 \pm 0.4$ & $0.11^{+0.06}_{-0.04}$ & 0.09 \\ 
26.7 & 50 & $2.9 \pm 0.3$ & $0.15^{+0.08}_{-0.05}$ & 0.12 \\ 
29.8 & 2 & $3.8 \pm 0.4$ & $0.12^{+0.06}_{-0.04}$ & 0.07 \\ 
29.8 & 5 & $4.5 \pm 0.4$ & $0.10^{+0.05}_{-0.03}$ & 0.06 \\ 
29.8 & 10 & $4.6 \pm 0.5$ & $0.10^{+0.05}_{-0.03}$ & 0.06 \\ 
29.8 & 25 & $3.9 \pm 0.5$ & $0.12^{+0.06}_{-0.04}$ & 0.07 \\ 
29.8 & 50 & $3.0 \pm 0.3$ & $0.15^{+0.08}_{-0.05}$ & 0.09 \\ 
33.3 & 2 & $3.7 \pm 0.4$ & $0.13^{+0.06}_{-0.04}$ & 0.15 \\ 
33.3 & 5 & $4.4 \pm 0.4$ & $0.11^{+0.05}_{-0.03}$ & 0.14 \\ 
33.3 & 10 & $4.5 \pm 0.5$ & $0.11^{+0.05}_{-0.03}$ & 0.14 \\ 
33.3 & 25 & $3.9 \pm 0.5$ & $0.12^{+0.06}_{-0.04}$ & 0.17 \\ 
33.3 & 50 & $3.0 \pm 0.3$ & $0.15^{+0.08}_{-0.05}$ & 0.20 \\ 
37.2 & 2 & $3.5 \pm 0.4$ & $0.14^{+0.07}_{-0.04}$ & 0.23 \\ 
37.2 & 5 & $4.2 \pm 0.4$ & $0.11^{+0.05}_{-0.04}$ & 0.20 \\ 
37.2 & 10 & $4.3 \pm 0.5$ & $0.11^{+0.06}_{-0.04}$ & 0.21 \\ 
37.2 & 25 & $3.8 \pm 0.5$ & $0.13^{+0.07}_{-0.04}$ & 0.24 \\ 
37.2 & 50 & $3.0 \pm 0.3$ & $0.16^{+0.08}_{-0.05}$ & 0.29 \\ 
41.7 & 2 & $3.4 \pm 0.4$ & $0.15^{+0.08}_{-0.05}$ & 0.16 \\ 
41.7 & 5 & $4.0 \pm 0.4$ & $0.12^{+0.06}_{-0.04}$ & 0.14 \\ 
41.7 & 10 & $4.1 \pm 0.5$ & $0.12^{+0.06}_{-0.04}$ & 0.15 \\ 
41.7 & 25 & $3.6 \pm 0.5$ & $0.14^{+0.07}_{-0.05}$ & 0.18 \\ 
41.7 & 50 & $2.9 \pm 0.3$ & $0.17^{+0.08}_{-0.05}$ & 0.21 \\ 
47.0 & 2 & $3.1 \pm 0.4$ & $0.16^{+0.08}_{-0.05}$ & 0.22 \\ 
47.0 & 5 & $3.6 \pm 0.4$ & $0.13^{+0.07}_{-0.04}$ & 0.20 \\ 
47.0 & 10 & $3.8 \pm 0.5$ & $0.13^{+0.07}_{-0.04}$ & 0.20 \\ 
47.0 & 25 & $3.4 \pm 0.5$ & $0.15^{+0.09}_{-0.05}$ & 0.25 \\ 
47.0 & 50 & $2.7 \pm 0.3$ & $0.18^{+0.09}_{-0.06}$ & 0.29 \\ 
50.0 & 2 & $2.8 \pm 0.4$ & $0.17^{+0.09}_{-0.06}$ & 0.19 \\ 
50.0 & 5 & $3.4 \pm 0.4$ & $0.14^{+0.07}_{-0.05}$ & 0.16 \\ 
50.0 & 10 & $3.6 \pm 0.5$ & $0.14^{+0.07}_{-0.04}$ & 0.17 \\ 
50.0 & 25 & $3.2 \pm 0.5$ & $0.16^{+0.09}_{-0.05}$ & 0.21 \\ 
50.0 & 50 & $2.5 \pm 0.3$ & $0.20^{+0.11}_{-0.07}$ & 0.26 \\ 
\bottomrule
\end{tabular}}
\end{minipage}
\label{tab:sup_dpp_eff_limit}
\end{table}

\begin{table}
\caption{Efficiency in percentage, and 95 $\%$ CL upper limits on the production cross-section times branching fraction for LLPs produced through the HIG mechanism.}
\renewcommand{\arraystretch}{1.12}
\begin{minipage}{.48\linewidth}
\resizebox{\textwidth}{!}{
\begin{tabular}{cccccc}
\toprule
\mLLP & \tauLLP & $\epsilon$ & expected UL & observed UL \\ 
\midrule
7.0 & 2 & $3.4 \pm 0.4$ & $0.19^{+0.10}_{-0.06}$ & 0.18 \\ 
7.0 & 5 & $3.6 \pm 0.5$ & $0.18^{+0.09}_{-0.06}$ & 0.16 \\ 
7.0 & 10 & $2.8 \pm 0.4$ & $0.23^{+0.13}_{-0.08}$ & 0.20 \\ 
7.0 & 25 & $1.5 \pm 0.3$ & $0.42^{+0.25}_{-0.15}$ & 0.36 \\ 
7.0 & 50 & $0.89 \pm 0.16$ & $0.76^{+0.42}_{-0.24}$ & 0.60 \\ 
7.8 & 2 & $3.4 \pm 0.4$ & $0.16^{+0.08}_{-0.05}$ & 0.19 \\ 
7.8 & 5 & $3.6 \pm 0.5$ & $0.15^{+0.08}_{-0.05}$ & 0.17 \\ 
7.8 & 10 & $2.9 \pm 0.5$ & $0.19^{+0.10}_{-0.06}$ & 0.21 \\ 
7.8 & 25 & $1.7 \pm 0.3$ & $0.34^{+0.19}_{-0.11}$ & 0.34 \\ 
7.8 & 50 & $0.98 \pm 0.17$ & $0.59^{+0.32}_{-0.19}$ & 0.57 \\ 
8.7 & 2 & $3.4 \pm 0.4$ & $0.14^{+0.07}_{-0.04}$ & 0.12 \\ 
8.7 & 5 & $3.7 \pm 0.5$ & $0.13^{+0.07}_{-0.04}$ & 0.11 \\ 
8.7 & 10 & $3.1 \pm 0.5$ & $0.16^{+0.08}_{-0.05}$ & 0.13 \\ 
8.7 & 25 & $1.8 \pm 0.3$ & $0.27^{+0.15}_{-0.09}$ & 0.21 \\ 
8.7 & 50 & $1.1 \pm 0.2$ & $0.47^{+0.25}_{-0.15}$ & 0.36 \\ 
9.7 & 2 & $3.4 \pm 0.4$ & $0.12^{+0.06}_{-0.04}$ & 0.08 \\ 
9.7 & 5 & $3.7 \pm 0.5$ & $0.11^{+0.06}_{-0.03}$ & 0.07 \\ 
9.7 & 10 & $3.2 \pm 0.5$ & $0.13^{+0.07}_{-0.04}$ & 0.08 \\ 
9.7 & 25 & $2.0 \pm 0.3$ & $0.21^{+0.12}_{-0.06}$ & 0.13 \\ 
9.7 & 50 & $1.2 \pm 0.2$ & $0.36^{+0.20}_{-0.11}$ & 0.21 \\ 
10.8 & 2 & $3.4 \pm 0.4$ & $0.10^{+0.05}_{-0.03}$ & 0.10 \\ 
10.8 & 5 & $3.8 \pm 0.5$ & $0.09^{+0.05}_{-0.03}$ & 0.10 \\ 
10.8 & 10 & $3.3 \pm 0.5$ & $0.10^{+0.06}_{-0.03}$ & 0.11 \\ 
10.8 & 25 & $2.1 \pm 0.3$ & $0.16^{+0.09}_{-0.05}$ & 0.16 \\ 
10.8 & 50 & $1.3 \pm 0.2$ & $0.27^{+0.15}_{-0.09}$ & 0.27 \\ 
12.1 & 2 & $3.4 \pm 0.4$ & $0.09^{+0.05}_{-0.03}$ & 0.07 \\ 
12.1 & 5 & $3.8 \pm 0.5$ & $0.08^{+0.04}_{-0.03}$ & 0.07 \\ 
12.1 & 10 & $3.4 \pm 0.5$ & $0.09^{+0.05}_{-0.03}$ & 0.07 \\ 
12.1 & 25 & $2.3 \pm 0.3$ & $0.14^{+0.08}_{-0.05}$ & 0.11 \\ 
12.1 & 50 & $1.4 \pm 0.2$ & $0.22^{+0.13}_{-0.08}$ & 0.17 \\ 
13.5 & 2 & $3.4 \pm 0.4$ & $0.09^{+0.05}_{-0.03}$ & 0.08 \\ 
13.5 & 5 & $3.9 \pm 0.5$ & $0.08^{+0.04}_{-0.03}$ & 0.07 \\ 
13.5 & 10 & $3.6 \pm 0.5$ & $0.08^{+0.05}_{-0.03}$ & 0.08 \\ 
13.5 & 25 & $2.4 \pm 0.4$ & $0.12^{+0.07}_{-0.04}$ & 0.11 \\ 
13.5 & 50 & $1.5 \pm 0.2$ & $0.20^{+0.11}_{-0.07}$ & 0.18 \\ 
15.1 & 2 & $3.4 \pm 0.4$ & $0.09^{+0.05}_{-0.03}$ & 0.15 \\ 
15.1 & 5 & $3.9 \pm 0.5$ & $0.08^{+0.04}_{-0.03}$ & 0.14 \\ 
15.1 & 10 & $3.7 \pm 0.5$ & $0.08^{+0.04}_{-0.03}$ & 0.15 \\ 
15.1 & 25 & $2.6 \pm 0.4$ & $0.12^{+0.07}_{-0.04}$ & 0.21 \\ 
15.1 & 50 & $1.6 \pm 0.3$ & $0.19^{+0.11}_{-0.07}$ & 0.32 \\ 
16.9 & 2 & $3.3 \pm 0.4$ & $0.10^{+0.05}_{-0.03}$ & 0.13 \\ 
16.9 & 5 & $3.9 \pm 0.5$ & $0.09^{+0.04}_{-0.03}$ & 0.11 \\ 
16.9 & 10 & $3.8 \pm 0.5$ & $0.09^{+0.05}_{-0.03}$ & 0.12 \\ 
16.9 & 25 & $2.7 \pm 0.4$ & $0.12^{+0.07}_{-0.04}$ & 0.16 \\ 
16.9 & 50 & $1.7 \pm 0.3$ & $0.19^{+0.11}_{-0.06}$ & 0.25 \\ 
18.9 & 2 & $3.3 \pm 0.4$ & $0.12^{+0.06}_{-0.04}$ & 0.08 \\ 
18.9 & 5 & $3.9 \pm 0.5$ & $0.10^{+0.05}_{-0.03}$ & 0.07 \\ 
18.9 & 10 & $3.8 \pm 0.6$ & $0.10^{+0.05}_{-0.03}$ & 0.07 \\ 
18.9 & 25 & $2.8 \pm 0.4$ & $0.14^{+0.07}_{-0.04}$ & 0.10 \\ 
18.9 & 50 & $1.8 \pm 0.3$ & $0.21^{+0.11}_{-0.07}$ & 0.15 \\ 
\bottomrule
\end{tabular}}
\end{minipage}
\hfill
\begin{minipage}{.48\linewidth}
\resizebox{\textwidth}{!}{
\begin{tabular}{cccccc}
\toprule
\mLLP & \tauLLP & $\epsilon$ & expected UL & observed UL \\ 
\midrule
21.2 & 2 & $3.3 \pm 0.4$ & $0.13^{+0.07}_{-0.04}$ & 0.09 \\ 
21.2 & 5 & $3.9 \pm 0.5$ & $0.11^{+0.06}_{-0.04}$ & 0.08 \\ 
21.2 & 10 & $3.9 \pm 0.6$ & $0.11^{+0.06}_{-0.04}$ & 0.08 \\ 
21.2 & 25 & $2.9 \pm 0.5$ & $0.15^{+0.08}_{-0.05}$ & 0.10 \\ 
21.2 & 50 & $1.9 \pm 0.3$ & $0.22^{+0.13}_{-0.08}$ & 0.15 \\ 
23.8 & 2 & $3.2 \pm 0.4$ & $0.14^{+0.08}_{-0.05}$ & 0.17 \\ 
23.8 & 5 & $3.8 \pm 0.5$ & $0.11^{+0.06}_{-0.04}$ & 0.16 \\ 
23.8 & 10 & $3.9 \pm 0.6$ & $0.11^{+0.07}_{-0.04}$ & 0.17 \\ 
23.8 & 25 & $3.0 \pm 0.5$ & $0.14^{+0.08}_{-0.05}$ & 0.20 \\ 
23.8 & 50 & $2.0 \pm 0.3$ & $0.21^{+0.12}_{-0.07}$ & 0.29 \\ 
26.7 & 2 & $3.1 \pm 0.4$ & $0.14^{+0.07}_{-0.05}$ & 0.10 \\ 
26.7 & 5 & $3.8 \pm 0.5$ & $0.12^{+0.06}_{-0.04}$ & 0.09 \\ 
26.7 & 10 & $3.8 \pm 0.6$ & $0.12^{+0.06}_{-0.04}$ & 0.10 \\ 
26.7 & 25 & $3.0 \pm 0.5$ & $0.15^{+0.08}_{-0.05}$ & 0.12 \\ 
26.7 & 50 & $2.1 \pm 0.3$ & $0.21^{+0.12}_{-0.07}$ & 0.17 \\ 
29.8 & 2 & $3.0 \pm 0.4$ & $0.15^{+0.08}_{-0.05}$ & 0.08 \\ 
29.8 & 5 & $3.7 \pm 0.5$ & $0.13^{+0.07}_{-0.04}$ & 0.07 \\ 
29.8 & 10 & $3.7 \pm 0.6$ & $0.13^{+0.07}_{-0.04}$ & 0.07 \\ 
29.8 & 25 & $3.0 \pm 0.5$ & $0.15^{+0.08}_{-0.05}$ & 0.09 \\ 
29.8 & 50 & $2.1 \pm 0.4$ & $0.21^{+0.12}_{-0.07}$ & 0.12 \\ 
33.3 & 2 & $2.9 \pm 0.4$ & $0.16^{+0.08}_{-0.05}$ & 0.19 \\ 
33.3 & 5 & $3.5 \pm 0.5$ & $0.14^{+0.07}_{-0.05}$ & 0.18 \\ 
33.3 & 10 & $3.6 \pm 0.6$ & $0.14^{+0.08}_{-0.05}$ & 0.18 \\ 
33.3 & 25 & $3.0 \pm 0.5$ & $0.16^{+0.09}_{-0.05}$ & 0.22 \\ 
33.3 & 50 & $2.2 \pm 0.4$ & $0.22^{+0.12}_{-0.07}$ & 0.28 \\ 
37.2 & 2 & $2.8 \pm 0.4$ & $0.17^{+0.09}_{-0.06}$ & 0.28 \\ 
37.2 & 5 & $3.4 \pm 0.5$ & $0.14^{+0.08}_{-0.05}$ & 0.26 \\ 
37.2 & 10 & $3.4 \pm 0.6$ & $0.14^{+0.08}_{-0.05}$ & 0.27 \\ 
37.2 & 25 & $2.9 \pm 0.6$ & $0.17^{+0.10}_{-0.05}$ & 0.32 \\ 
37.2 & 50 & $2.2 \pm 0.4$ & $0.21^{+0.12}_{-0.07}$ & 0.39 \\ 
41.7 & 2 & $2.7 \pm 0.4$ & $0.19^{+0.09}_{-0.06}$ & 0.20 \\ 
41.7 & 5 & $3.2 \pm 0.5$ & $0.16^{+0.09}_{-0.05}$ & 0.19 \\ 
41.7 & 10 & $3.3 \pm 0.6$ & $0.16^{+0.09}_{-0.05}$ & 0.19 \\ 
41.7 & 25 & $2.8 \pm 0.6$ & $0.18^{+0.11}_{-0.06}$ & 0.23 \\ 
41.7 & 50 & $2.2 \pm 0.4$ & $0.22^{+0.12}_{-0.07}$ & 0.28 \\ 
47.0 & 2 & $2.6 \pm 0.3$ & $0.19^{+0.10}_{-0.06}$ & 0.27 \\ 
47.0 & 5 & $3.1 \pm 0.5$ & $0.16^{+0.09}_{-0.05}$ & 0.24 \\ 
47.0 & 10 & $3.2 \pm 0.6$ & $0.16^{+0.09}_{-0.05}$ & 0.26 \\ 
47.0 & 25 & $2.8 \pm 0.6$ & $0.18^{+0.11}_{-0.06}$ & 0.31 \\ 
47.0 & 50 & $2.2 \pm 0.4$ & $0.22^{+0.12}_{-0.07}$ & 0.35 \\ 
50.0 & 2 & $2.5 \pm 0.3$ & $0.20^{+0.10}_{-0.06}$ & 0.21 \\ 
50.0 & 5 & $3.1 \pm 0.5$ & $0.16^{+0.09}_{-0.05}$ & 0.19 \\ 
50.0 & 10 & $3.2 \pm 0.6$ & $0.16^{+0.09}_{-0.05}$ & 0.20 \\ 
50.0 & 25 & $2.8 \pm 0.6$ & $0.18^{+0.11}_{-0.06}$ & 0.24 \\ 
50.0 & 50 & $2.3 \pm 0.4$ & $0.22^{+0.12}_{-0.07}$ & 0.29 \\ 
\bottomrule
\end{tabular}}
\end{minipage}
\label{tab:sup_hig_eff_limit}
\end{table}

\begin{table}
\caption{Efficiency in percentage, and 95 $\%$ CL upper limits on the production cross-section times branching fraction for LLPs produced through the CC mechanism.}
\renewcommand{\arraystretch}{1.12}
\begin{minipage}{.48\linewidth}
\resizebox{\textwidth}{!}{
\begin{tabular}{cccccc}
\toprule
\mLLP & \tauLLP & $\epsilon$ & expected UL & observed UL \\ 
\midrule
7.0 & 2 & $1.5 \pm 0.2$ & $0.48^{+0.24}_{-0.16}$ & 0.45 \\ 
7.0 & 5 & $1.6 \pm 0.2$ & $0.42^{+0.21}_{-0.14}$ & 0.38 \\ 
7.0 & 10 & $1.4 \pm 0.2$ & $0.51^{+0.26}_{-0.17}$ & 0.44 \\ 
7.0 & 25 & $0.85 \pm 0.13$ & $0.83^{+0.46}_{-0.27}$ & 0.73 \\ 
7.0 & 50 & $0.50 \pm 0.09$ & $1.52^{+0.84}_{-0.48}$ & 1.22 \\ 
7.8 & 2 & $1.5 \pm 0.2$ & $0.41^{+0.20}_{-0.13}$ & 0.50 \\ 
7.8 & 5 & $1.7 \pm 0.2$ & $0.36^{+0.17}_{-0.11}$ & 0.42 \\ 
7.8 & 10 & $1.5 \pm 0.2$ & $0.42^{+0.21}_{-0.13}$ & 0.46 \\ 
7.8 & 25 & $0.90 \pm 0.14$ & $0.70^{+0.36}_{-0.22}$ & 0.73 \\ 
7.8 & 50 & $0.54 \pm 0.09$ & $1.18^{+0.63}_{-0.38}$ & 1.19 \\ 
8.7 & 2 & $1.5 \pm 0.2$ & $0.35^{+0.17}_{-0.11}$ & 0.30 \\ 
8.7 & 5 & $1.7 \pm 0.2$ & $0.31^{+0.15}_{-0.09}$ & 0.26 \\ 
8.7 & 10 & $1.5 \pm 0.2$ & $0.35^{+0.17}_{-0.11}$ & 0.29 \\ 
8.7 & 25 & $0.96 \pm 0.14$ & $0.56^{+0.29}_{-0.17}$ & 0.43 \\ 
8.7 & 50 & $0.58 \pm 0.09$ & $0.95^{+0.49}_{-0.30}$ & 0.72 \\ 
9.7 & 2 & $1.5 \pm 0.2$ & $0.30^{+0.15}_{-0.09}$ & 0.19 \\ 
9.7 & 5 & $1.7 \pm 0.2$ & $0.26^{+0.13}_{-0.08}$ & 0.16 \\ 
9.7 & 10 & $1.5 \pm 0.2$ & $0.28^{+0.15}_{-0.09}$ & 0.17 \\ 
9.7 & 25 & $1.0 \pm 0.1$ & $0.44^{+0.23}_{-0.13}$ & 0.26 \\ 
9.7 & 50 & $0.63 \pm 0.09$ & $0.73^{+0.39}_{-0.22}$ & 0.43 \\ 
10.8 & 2 & $1.5 \pm 0.2$ & $0.26^{+0.13}_{-0.08}$ & 0.26 \\ 
10.8 & 5 & $1.7 \pm 0.2$ & $0.22^{+0.11}_{-0.07}$ & 0.22 \\ 
10.8 & 10 & $1.6 \pm 0.2$ & $0.23^{+0.12}_{-0.07}$ & 0.23 \\ 
10.8 & 25 & $1.1 \pm 0.1$ & $0.34^{+0.18}_{-0.11}$ & 0.34 \\ 
10.8 & 50 & $0.68 \pm 0.10$ & $0.55^{+0.29}_{-0.18}$ & 0.53 \\ 
12.1 & 2 & $1.5 \pm 0.2$ & $0.22^{+0.12}_{-0.08}$ & 0.19 \\ 
12.1 & 5 & $1.7 \pm 0.2$ & $0.19^{+0.10}_{-0.06}$ & 0.16 \\ 
12.1 & 10 & $1.6 \pm 0.2$ & $0.20^{+0.11}_{-0.07}$ & 0.17 \\ 
12.1 & 25 & $1.2 \pm 0.2$ & $0.29^{+0.16}_{-0.10}$ & 0.23 \\ 
12.1 & 50 & $0.74 \pm 0.10$ & $0.45^{+0.25}_{-0.15}$ & 0.36 \\ 
13.5 & 2 & $1.5 \pm 0.1$ & $0.22^{+0.12}_{-0.07}$ & 0.21 \\ 
13.5 & 5 & $1.8 \pm 0.2$ & $0.18^{+0.10}_{-0.06}$ & 0.18 \\ 
13.5 & 10 & $1.7 \pm 0.2$ & $0.18^{+0.10}_{-0.06}$ & 0.18 \\ 
13.5 & 25 & $1.2 \pm 0.2$ & $0.26^{+0.14}_{-0.09}$ & 0.24 \\ 
13.5 & 50 & $0.80 \pm 0.11$ & $0.41^{+0.22}_{-0.14}$ & 0.37 \\ 
15.1 & 2 & $1.5 \pm 0.1$ & $0.22^{+0.11}_{-0.07}$ & 0.38 \\ 
15.1 & 5 & $1.8 \pm 0.2$ & $0.18^{+0.10}_{-0.06}$ & 0.33 \\ 
15.1 & 10 & $1.7 \pm 0.2$ & $0.19^{+0.10}_{-0.06}$ & 0.33 \\ 
15.1 & 25 & $1.3 \pm 0.2$ & $0.25^{+0.14}_{-0.09}$ & 0.44 \\ 
15.1 & 50 & $0.86 \pm 0.11$ & $0.39^{+0.21}_{-0.13}$ & 0.67 \\ 
16.9 & 2 & $1.5 \pm 0.1$ & $0.24^{+0.12}_{-0.08}$ & 0.32 \\ 
16.9 & 5 & $1.8 \pm 0.2$ & $0.20^{+0.10}_{-0.06}$ & 0.27 \\ 
16.9 & 10 & $1.8 \pm 0.2$ & $0.20^{+0.10}_{-0.06}$ & 0.27 \\ 
16.9 & 25 & $1.4 \pm 0.2$ & $0.26^{+0.14}_{-0.08}$ & 0.34 \\ 
16.9 & 50 & $0.92 \pm 0.12$ & $0.40^{+0.21}_{-0.12}$ & 0.51 \\ 
18.9 & 2 & $1.5 \pm 0.2$ & $0.28^{+0.13}_{-0.09}$ & 0.19 \\ 
18.9 & 5 & $1.9 \pm 0.2$ & $0.23^{+0.11}_{-0.07}$ & 0.16 \\ 
18.9 & 10 & $1.8 \pm 0.2$ & $0.23^{+0.11}_{-0.07}$ & 0.17 \\ 
18.9 & 25 & $1.4 \pm 0.2$ & $0.29^{+0.14}_{-0.09}$ & 0.20 \\ 
18.9 & 50 & $0.98 \pm 0.13$ & $0.43^{+0.22}_{-0.14}$ & 0.29 \\ 
\bottomrule
\end{tabular}}
\end{minipage}
\hfill
\begin{minipage}{.48\linewidth}
\resizebox{\textwidth}{!}{
\begin{tabular}{cccccc}
\toprule
\mLLP & \tauLLP & $\epsilon$ & expected UL & observed UL \\ 
\midrule
21.2 & 2 & $1.5 \pm 0.2$ & $0.32^{+0.16}_{-0.10}$ & 0.21 \\ 
21.2 & 5 & $1.9 \pm 0.2$ & $0.25^{+0.13}_{-0.08}$ & 0.17 \\ 
21.2 & 10 & $1.9 \pm 0.2$ & $0.25^{+0.13}_{-0.08}$ & 0.18 \\ 
21.2 & 25 & $1.5 \pm 0.2$ & $0.31^{+0.16}_{-0.10}$ & 0.21 \\ 
21.2 & 50 & $1.0 \pm 0.1$ & $0.45^{+0.24}_{-0.15}$ & 0.30 \\ 
23.8 & 2 & $1.5 \pm 0.2$ & $0.31^{+0.17}_{-0.10}$ & 0.41 \\ 
23.8 & 5 & $1.9 \pm 0.2$ & $0.24^{+0.13}_{-0.08}$ & 0.36 \\ 
23.8 & 10 & $1.9 \pm 0.2$ & $0.24^{+0.13}_{-0.08}$ & 0.37 \\ 
23.8 & 25 & $1.6 \pm 0.2$ & $0.29^{+0.16}_{-0.10}$ & 0.42 \\ 
23.8 & 50 & $1.1 \pm 0.1$ & $0.41^{+0.23}_{-0.14}$ & 0.57 \\ 
26.7 & 2 & $1.5 \pm 0.2$ & $0.30^{+0.16}_{-0.10}$ & 0.22 \\ 
26.7 & 5 & $1.9 \pm 0.2$ & $0.24^{+0.12}_{-0.08}$ & 0.20 \\ 
26.7 & 10 & $2.0 \pm 0.2$ & $0.24^{+0.13}_{-0.08}$ & 0.20 \\ 
26.7 & 25 & $1.6 \pm 0.2$ & $0.28^{+0.15}_{-0.09}$ & 0.24 \\ 
26.7 & 50 & $1.2 \pm 0.2$ & $0.40^{+0.21}_{-0.13}$ & 0.32 \\ 
29.8 & 2 & $1.6 \pm 0.2$ & $0.30^{+0.15}_{-0.10}$ & 0.16 \\ 
29.8 & 5 & $2.0 \pm 0.2$ & $0.24^{+0.12}_{-0.08}$ & 0.14 \\ 
29.8 & 10 & $2.0 \pm 0.2$ & $0.24^{+0.12}_{-0.08}$ & 0.15 \\ 
29.8 & 25 & $1.7 \pm 0.2$ & $0.28^{+0.15}_{-0.09}$ & 0.17 \\ 
29.8 & 50 & $1.2 \pm 0.2$ & $0.38^{+0.20}_{-0.13}$ & 0.22 \\ 
33.3 & 2 & $1.6 \pm 0.2$ & $0.30^{+0.15}_{-0.10}$ & 0.35 \\ 
33.3 & 5 & $2.0 \pm 0.2$ & $0.25^{+0.12}_{-0.08}$ & 0.31 \\ 
33.3 & 10 & $2.0 \pm 0.2$ & $0.24^{+0.12}_{-0.08}$ & 0.32 \\ 
33.3 & 25 & $1.7 \pm 0.2$ & $0.29^{+0.15}_{-0.09}$ & 0.38 \\ 
33.3 & 50 & $1.3 \pm 0.2$ & $0.37^{+0.19}_{-0.12}$ & 0.49 \\ 
37.2 & 2 & $1.6 \pm 0.2$ & $0.30^{+0.15}_{-0.09}$ & 0.47 \\ 
37.2 & 5 & $2.0 \pm 0.2$ & $0.24^{+0.12}_{-0.08}$ & 0.42 \\ 
37.2 & 10 & $2.0 \pm 0.2$ & $0.24^{+0.12}_{-0.08}$ & 0.43 \\ 
37.2 & 25 & $1.7 \pm 0.2$ & $0.27^{+0.14}_{-0.09}$ & 0.52 \\ 
37.2 & 50 & $1.3 \pm 0.2$ & $0.36^{+0.19}_{-0.12}$ & 0.65 \\ 
41.7 & 2 & $1.7 \pm 0.2$ & $0.30^{+0.15}_{-0.09}$ & 0.32 \\ 
41.7 & 5 & $2.0 \pm 0.2$ & $0.24^{+0.12}_{-0.08}$ & 0.28 \\ 
41.7 & 10 & $2.1 \pm 0.3$ & $0.23^{+0.12}_{-0.08}$ & 0.29 \\ 
41.7 & 25 & $1.8 \pm 0.2$ & $0.28^{+0.15}_{-0.09}$ & 0.36 \\ 
41.7 & 50 & $1.4 \pm 0.2$ & $0.35^{+0.18}_{-0.12}$ & 0.44 \\ 
47.0 & 2 & $1.7 \pm 0.2$ & $0.29^{+0.14}_{-0.10}$ & 0.40 \\ 
47.0 & 5 & $2.1 \pm 0.2$ & $0.23^{+0.11}_{-0.07}$ & 0.35 \\ 
47.0 & 10 & $2.1 \pm 0.3$ & $0.22^{+0.11}_{-0.07}$ & 0.35 \\ 
47.0 & 25 & $1.8 \pm 0.2$ & $0.26^{+0.15}_{-0.09}$ & 0.44 \\ 
47.0 & 50 & $1.4 \pm 0.2$ & $0.33^{+0.18}_{-0.11}$ & 0.53 \\ 
50.0 & 2 & $1.7 \pm 0.2$ & $0.29^{+0.14}_{-0.09}$ & 0.31 \\ 
50.0 & 5 & $2.1 \pm 0.2$ & $0.22^{+0.11}_{-0.07}$ & 0.26 \\ 
50.0 & 10 & $2.2 \pm 0.3$ & $0.22^{+0.11}_{-0.07}$ & 0.27 \\ 
50.0 & 25 & $1.9 \pm 0.2$ & $0.26^{+0.13}_{-0.08}$ & 0.34 \\ 
50.0 & 50 & $1.5 \pm 0.2$ & $0.34^{+0.17}_{-0.11}$ & 0.43 \\ 
\bottomrule
\end{tabular}}
\end{minipage}
\label{tab:sup_cc_eff_limit}
\end{table}